\documentclass[aps,prd,showpacs,twocolumn,superscriptaddress,sort&compress,floatfix]{revtex4-1}

\usepackage{graphicx}
\usepackage{amsmath}

\usepackage{microtype}

\begin{document}

\title{Effects of a dressed quark-gluon vertex in pseudoscalar heavy-light mesons}


\author{M. G\'{o}mez-Rocha}
\email[]{maria.gomez-rocha@uni-graz.at}
\affiliation{Institute of High Energy Physics, Austrian Academy of Sciences, A-1050 Vienna, Austria}
\affiliation{Institute of Physics, University of Graz, NAWI Graz, A-8010 Graz, Austria}

\author{T. Hilger}
\email[]{thomas.hilger@uni-graz.at}
\affiliation{Institute of Physics, University of Graz, NAWI Graz, A-8010 Graz, Austria}

\author{A. Krassnigg}
\email[]{andreas.krassnigg@uni-graz.at}
\affiliation{Institute of Physics, University of Graz, NAWI Graz, A-8010 Graz, Austria}

\date{\today}

\begin{abstract}
Using a simple model in the context of the Dyson-Schwinger-Bethe-Salpeter
approach, we investigate the effects of a dressed-quark-gluon vertex on
pseudoscalar meson masses. In particular, we focus on the unequal-mass case
and investigate heavy-light meson masses; in addition, we study the premise of 
the effective treatment of heavy quarks in our approach.

\end{abstract}


\maketitle

\section{ Introduction }\label{sec:intro}

The realm of heavy-light mesons is topical and promising for a number of 
reasons. In the light of quantum chromodynamics (QCD), one faces the
challenge of a multi-scale problem, where many successful theoretical \emph{Ans\"atze}
can be tested, compared, and brought together. With the chiral limit on one and 
the heavy-quark regime on the other end of the quark mass range, an ideal approach
should respect the symmetries apparent in QCD at both of these ends together with
the other basic properties of the theory, e.\,g., its perturbative limit, and
a realization of confinement. Overall, a nonperturbative technique is necessary
and has the benefits of wide applicability, if used properly and with a model
tailored to this task.

In particular, it has been shown in \cite{Hilger:2009kn,Hilger:2010zb,Hilger:2010cn,%
Hilger:2011cq,Hilger:2012db,Buchheim:2014rpa} that the spectral difference 
of parity partners in the sector of mesons with a heavy and a light valence quark 
is driven by a subtle balance between a suppressed interaction with the QCD ground 
state and an enhanced impact of light-quark chiral-symmetry breaking effects by the 
heavy-quark mass. The mass splitting of these parity partners is comparable to the 
ones in the sector of mesons with light valence quarks only, making them a suitable 
object for related investigations of dynamical chiral-symmetry breaking.

The Dyson-Schwinger-Bethe-Salpeter-equation (DSBSE) approach, a modern nonperturbative
tool for quantum field theory \cite{Roberts:2007jh,Fischer:2006ub,Alkofer:2000wg,Sanchis-Alepuz:2015tha} 
complementary to the well-known lattice-regularized 
approach \cite{Dudek:2007wv,Liu:2012ze,Thomas:2014dpa,Flynn:2015mha}, 
is an excellent candidate for such a study, since all requirements are met.
In this work we build on an earlier investigation of a systematic approach to 
dressing the quark-gluon vertex (QGV) and thus, consistently the quark Dyson-Schwinger-equation 
(DSE) and meson Bethe-Salpeter-equation (BSE) integration-equation kernels 
\cite{Bhagwat:2004hn,Gomez-Rocha:2014vsa}, which are the necessary prerequisites for a meson study. 
Vertex dressing affects the BSE kernel in the various mesonic $J^{PC}$ channels differently,
which is in accord with expectations from the quark model, where different terms in the
potential also show varying importance in different channels, see, e.\,g., \cite{Godfrey:1985xj}.

In modern DSBSE studies with sophisticated effective interactions, a simple truncation 
offers the possibilities of comprehensive investigations, see
\cite{Maris:1999nt,Holl:2004fr,Holl:2005vu,Krassnigg:2009zh,Krassnigg:2010mh,Dorkin:2010ut,Mader:2011zf,%
UweHilger:2012uua,Popovici:2014pha,Hilger:2014nma,Fischer:2014xha,Fischer:2014cfa,Hilger:2015hka} and references therein.
Beyond the most popular rainbow-ladder (RL) truncation, there have also been sophisticated studies, 
following systematic schemata based on the structure of the DSE tower. However,
the numerical effort \cite{Bhagwat:2007rj,Krassnigg:2008gd,Blank:2010bp,Blank:2010sn,Blank:2011qk} quickly increases 
\cite{Watson:2004jq,Watson:2004kd,Fischer:2005en,Fischer:2008wy,Fischer:2009jm,Williams:2009wx,%
Williams:2014iea,Sanchis-Alepuz:2014wea,Chang:2009zb,Heupel:2014ina,Sanchis-Alepuz:2015qra} 
and so there have been a number of investigations,
like our present one, which make use of a very simple effective interaction \cite{Munczek:1983dx} in order
to be able to highlight particular schemes or effects, if one is able to sum up certain
classes of diagrams or study a certain scheme in as great detail as possible 
\cite{Bender:1996bb,Alkofer:2000wg,Bender:2002as,Krassnigg:2003dr,Bhagwat:2004hn,Holl:2004qn,%
Matevosyan:2006bk,Matevosyan:2007cx,Jinno:2015sea}. Apart from elucidating effects of high-order 
dressing terms or fully summed subclasses of
diagrams, such studies provide a guide as to how far a systematic truncation scheme
with a sophisticated effective interaction would have to go in order to achieve 
reliable results for various hadron properties.

Our focus on heavy-light meson systems can help to study the importance of QGV
correction terms by testing those in a multiply unbalanced system of one light (and at the same time
very strongly-dressed) antiquark together with one heavy (and only mildly-dressed) quark in a 
relativistic bound state. Note that investigations of heavy-quark propagators have shown that dressing
can have sizeable effects even for the case of b-quarks \cite{Nguyen:2009if,Souchlas:2010zz,Nguyen:2010yh}. In contrast
to earlier DSBSE studies making use of particular assumptions about heavy-quark propagators
\cite{Ivanov:1997iu,Ivanov:1997yg,Ivanov:1998ms,Blaschke:2000zm,Bhagwat:2006xi,Ivanov:2007cw,ElBennich:2009vx,ElBennich:2010ha}
we would like to motivate or justify these assumptions rather than start out from them.
Ultimately, we aim at a direct check of heavy-quark symmetry predictions \cite{Neubert:1993mb} like the ones performed recently,
e.\,g., in relativistic Hamiltonian dynamics \cite{GomezRocha:2012zd}.

The article is organized as follows: In Sec.~\ref{sec:setup} we present the setup used for the quark
DSE, the QGV, and the meson BSE. Results are collected and discussed in Sec.~\ref{sec:results} for
both the quark propagator dressing functions and the meson masses. After the conclusions in 
Sec.~\ref{sec:conclusions} we have collected more technical details and explanations as well as
a complete set of results in terms of tables and figures together with an analysis of
heavy-quark effective quark propagators and their comparison to the ones computed here in several appendices.

\section{Setup}\label{sec:setup}

Since this work is an extension of \cite{Bender:2002as,Bhagwat:2004hn}, we focus on essentials and differences
instead of repeating every detail here. Our calculations are performed in Euclidean space.

\subsection{Quark DSE}

For the purpose of investigating hadrons as bound states of (anti)quarks interacting via gluons, 
the DSE for the quark propagator is an adequate starting point, because it contains two main
ingredients of such a study, namely the renormalized dressed gluon propagator $D_{\mu\nu}$
and the renormalized dressed QGV $\Gamma^a_\nu$ with the color index ${}^a$. Its solution is the 
renomalized dressed quark propagator $S$ for a particular quark flavor, which has the general structure
\begin{eqnarray} 
\label{eq:quarkprop} 
S(p)^{-1}&=&i\gamma\cdot p A(p^2)+B(p^2)\\
&=&A(p^2)\left(i\gamma\cdot p + M(p^2) \right)\;.
\end{eqnarray} 

In this and the following, we omit dependences on renormalization and regularization 
parameters and details due to the particular ultraviolet-finite model interaction we 
are using herein: the corresponding renormalization constants are $=1$. 

The dressing functions $A$ and $B$ characterize $S$ completely. Alternatively,
one can use $A$ and $M$, all of which are functions of the quark momentum $p$ squared,
and implicitly also of the current-quark mass $m_q$ and thus the quark's flavor.
We write $s:=p^2$ and choose to investigate $A(s)$ and $M(s)$ below.

The quark DSE reads
\begin{eqnarray}\label{eq:selfenergy}
 S^{-1}(p)&=& S^{-1}_0 + \Sigma(p)=i \gamma\cdot p+m_q + \Sigma(p)\\\label{eq:fullquarkdse}
 &=& S^{-1}_0+ \int_q \!\!g^2 
 D_{\mu\nu}(p-q)\frac{\lambda^a}{2}\gamma_\mu S(q)\Gamma^a_\nu(q;p)
\end{eqnarray}
with $\int_q:=\int\frac{d^4q}{(2\pi)^4}$, $\Sigma$ the quark self energy, $p$ the quark momentum,
and $g$ the strong coupling constant. The subscript ${}_0$ denotes a bare quantity. With the specification of the effective 
interaction with constant strength $\mathcal{G}$ to be used herein \cite{Munczek:1983dx} via
\begin{equation} \label{eq:mnmodel} 
g^2 \, D_{\mu\nu}(k) := 
\left(\delta_{\mu\nu} - \frac{k_\mu k_\nu}{k^2}\right) (2\pi)^4\, \mathcal{G}^2 \, 
\delta^4(k)\,,
\end{equation} 
one arrives at an algebraic equation for $S$; the same is also true for the other equations
of relevance herein, namely the DSE for the QGV and the meson BSE. More
precisely, with the explicit color prescription $\Gamma_\mu^a(p)=\frac{\lambda^a}{2} \Gamma_\mu(p)$
and setting $\mathcal{G}=1$, thereby obtaining all dimensioned quantities in appropriate units 
of $\mathcal{G}$ one obtains
\begin{equation}\label{eq:quarkdse}
 S^{-1}(p)= i \gamma\cdot p+m_q + \gamma_\mu S(p)\Gamma_\nu^\mathcal{C}(p)\;,
\end{equation}
where the model parameter $\mathcal{C}$ introduced at this point is defined below. At this point
the quark DSE can be solved for any explicit form of the QGV, and results in physical units are 
obtained by providing the necessary input, i.\,e., a value for the model strength parameter $\mathcal{G}$ 
and a quark mass  $m_q$ for any given quark flavor.

To proceed, one can make use of the DSE for the QGV in order to define a model for $\Gamma_\nu^\mathcal{C}(p)$.
In particular, again following \cite{Bhagwat:2004hn}, the QGV DSE is used to obtain the effective equation 
\begin{equation}\label{eq:qgvdse}
\Gamma_\mu^\mathcal{C}(p)=\gamma_\mu-\mathcal{C}\,\gamma_\rho\, S(p)\,\Gamma_\mu^\mathcal{C}(p)\,S(p)\,\gamma_\rho\;,
\end{equation}
whose dependence on $\mathcal{C}$ stems from the effective combination of the abelian and non-abelian correction terms
in the QGV DSE such that: i) any differences in strength or momentum dependence of these two terms is taken to be proportional
to the simple and herein accessible structure of the abelian term; ii) the sign is that of the non-abelian correction
due to its dominant role; and iii) $\mathcal{C}$ is chosen in accordance with either lattice QCD, phenomenology, or
another reasonable point of comparison.

From a general point of view, the following values are of interest: $\mathcal{C}=-1/8$ corresponds to the case 
with abelian-only dressing as, e.\,g., used in \cite{Bender:2002as}. $\mathcal{C}=0$ corresponds to the popular 
RL truncation, which is also the zeroth order in the systematic schemes of the kind considered here. 
Finally, $\mathcal{C}=0.51$ was used in \cite{Bhagwat:2004hn}
as a result from fitting to lattice quark propagators, with good phenomenological success. 
We fix $\mathcal{C}=0.51$ throughout herein in order to avoid the overwhelming amounts of data that would result
from a thorough investigation of variation in all available parameters. For our purposes it is best to continue
from \cite{Bhagwat:2004hn} for clarity of the resulting effects, easy comparison, and direct applicability.

Equation (\ref{eq:qgvdse}) is the immediate basis for an iterative prescription, where the bare QGV serves 
as a starting value: 
$\Gamma_{\mu,0}^\mathcal{C}(p)=\gamma_\mu$. The recursion relation is
\begin{equation}\label{eq:qgvrecursion}
\Gamma_{\mu,i}^\mathcal{C}(p)=-\mathcal{C}\,\gamma_\rho\, S(p)\,\Gamma_{\mu,i-1}^\mathcal{C}(p)\,S(p)\,\gamma_\rho
\end{equation}
so that at a given order $n$ in this scheme one has for the QGV
\begin{equation}\label{eq:summedvertex}
\Gamma_\mu^\mathcal{C}(p)=\sum_{i=0}^n \Gamma_{\mu,i}^\mathcal{C}(p).
\end{equation}
and the final result for the QGV is obtained by $n\rightarrow\infty$.

At this point it is also important to note that, like the solution of the quark DSE, also the corresponding
result for the QGV implicitly depends on the flavor (and $m_q$) of the quark it is associated with, i.\,e.,
the details and properties of the factors $S(p)$ in Eqs.~(\ref{eq:qgvdse}) and (\ref{eq:qgvrecursion}). Note
also that, since the quark-gluon interaction is not flavor changing, both factors $S(p)$ in each of these
equations must correspond to the same flavor and value of $m_q$.

\subsection{Meson BSE}\label{sec:bse}

The meson BSE reads
\begin{equation} \label{eq:bse} 
[\Gamma(P;k)]_{EF} =\int_q [ K(P;k;q)]_{EF}^{GH} [S(q_+)\, \Gamma(P;q) \,S(q_-)]_{GH} 
\end{equation} 
where the four-vector arguments $P$ and $k$, $q$ are the quark-antiquark pair's total and relative momenta, respectively. 
$E, F,G ,H$ stand for color, flavor and spinor indices, $\Gamma(P;q)$ is the meson's Bethe-Salpeter amplitude (BSA),
and $K$ the fully-amputated dressed-quark-antiquark scattering kernel. The combination of quark propagators and
the BSA is often combined to the so-called Bethe-Salpeter wave function $\chi$ and one writes
\begin{equation} \label{eq:chi} 
\chi(P;q): =S(q_+)\, \Gamma(P;q) \,S(q_-)\;, 
\end{equation} 
where the meson flavor is determined by the combination of quark flavors from the two factors of $S$.
The quark and antiquark momenta are defined as $q_+=q+\eta P$ and $q_-=q-(1-\eta)P$. $\eta$ is called the momentum
partitioning parameter, and its arbitrariness in any computation is equivalent to the freedom in the definition
of the quark-antiquark relative momentum. In a covariant setup, any observable should be independent of $\eta$; 
still, numerical approximations, truncations, or other model defects can destroy this independence, as is the
case here, where the BSA is incomplete due to the particular form of the model interaction in Eq.~(\ref{eq:mnmodel}).
As a result, any study using this particular interaction should investigate also the $\eta$-dependence of
numerical results; such an analysis was also performed in Ref.~\cite{Munczek:1983dx} where this particular
form of effective interaction was introduced in RL truncation. Still, such
a model artifact does not destroy the model's capacity to elucidate our targeted meson properties and can
be easily quantified; thus it is well under control.

Provided one has solutions of the quark DSE for a given form of the quark
self-energy $\Sigma$, taking into account a particular form $\Gamma_\nu$ of the QGV, one can construct a 
BSE interaction kernel consistent with the quark propagator dressing functions via the axial-vector Ward-Takahashi 
identity (AVWTI) \cite{Munczek:1994zz,Maris:1997tm}. It was shown in \cite{Bender:1996bb,Bender:2002as,Bhagwat:2004hn} 
that the BSE kernel corresponding to our particular setup when omitting contributions
from gluon-unquenching \cite{Watson:2004jq,Fischer:2005en} for equal-mass constituents in the BSE reads
\begin{eqnarray}\nonumber 
\Gamma^M(P;k) &=& -\frac{4}{3} \int_q D_{\mu\nu}(k - q)\left[ \gamma_\mu \chi^M(P;q)\,\Gamma_\nu(q_-,k_-)  \right.\\
&+& \left.  \gamma_\mu  S(q_+)  \, \Lambda^M_\nu(P;k;q) \right]\;. \label{eq:origbsekernel}
\end{eqnarray} 
The color trace has been carried out at this point, since the mesonic case is straight-forward in this regard; 
in the following, we always give results with evaluated color traces. 
The superscript label ${}^M$ denotes applicability to a particular meson, e.\,g., a pseudoscalar, which we
investigate herein. This is important, because the structure of the correction term $\Lambda^M_\nu$ crucially depends on the
structure of the corresponding BSA. We will detail this point below for the pseudoscalar case.

While the first term on the r.h.s.\ of Eq.~(\ref{eq:origbsekernel}) is constructed immediately from the
given QGV, the construction of the second term is again based on a recursion relation analogous to the
one for the QGV. One sums up the correction terms up to a particular order $n$ to get $\Lambda^M$: 
\begin{equation}\label{eq:lambda}
\Lambda^M_\nu(P;k;q) = \sum_{i=0}^n \Lambda_{\nu,i}^M(P;k;q)\,.
\end{equation} 
The full result is then obtained by $n\rightarrow\infty$.

Generalizing to the unequal-mass case, we symmetrize Eq.~(\ref{eq:origbsekernel}) to get
\begin{eqnarray}\nonumber 
\Gamma^M(P;k) &=& -\frac{1}{2}\frac{4}{3}\int_q D_{\mu\nu}(k - q)\left[ \gamma_\mu \chi^M(P;q)\,\Gamma_\nu(q_-,k_-)  \right.\\ \nonumber
&+&  \gamma_\mu  S(q_+)  \, \Lambda^M_\nu(P;k;q) +  \Gamma_\mu(q_+,k_+)\chi^M(P;q)\gamma_\nu\\
&+& \left. \Lambda^M_\mu(P;k;q)  S(q_-)  \gamma_\nu  \right]\;. \label{eq:genbsekernel}
\end{eqnarray} 
The flavor content of this equation is encoded via the quark propagators and, more precisely, their arguments, 
the quark momenta $k_\pm, q_\pm$: any factor of $S$ or $\Gamma_\nu$ with (an) argument(s) with subscript ${}_+$ 
corresponds to the quark $1$ with flavor $1$ and those with (an) argument(s) with subscript ${}_-$ 
corresponds to the antiquark $2$ with flavor $2$. We note here also that in our setup for unequal
quark masses, the heavier one is associated with the subscript ${}_+$.
 
\begin{figure*}[t]
  \includegraphics[width=\columnwidth]{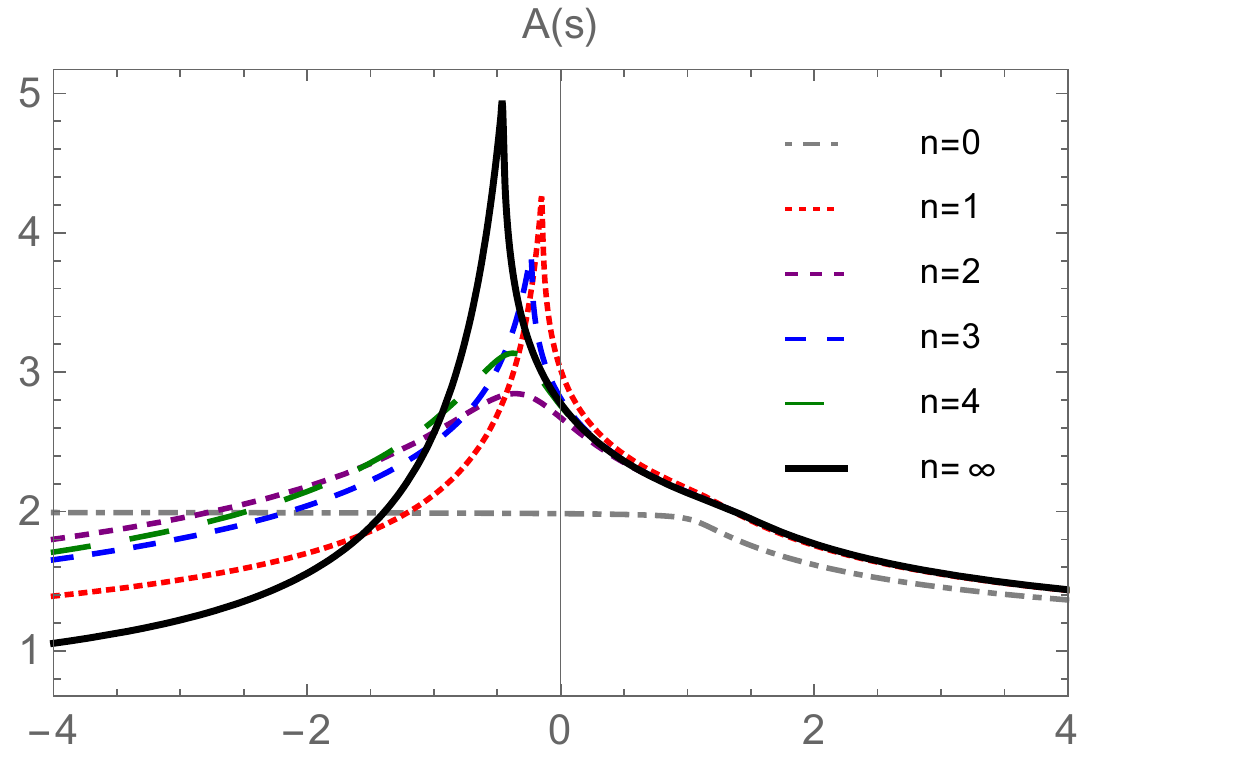}
  \includegraphics[width=\columnwidth]{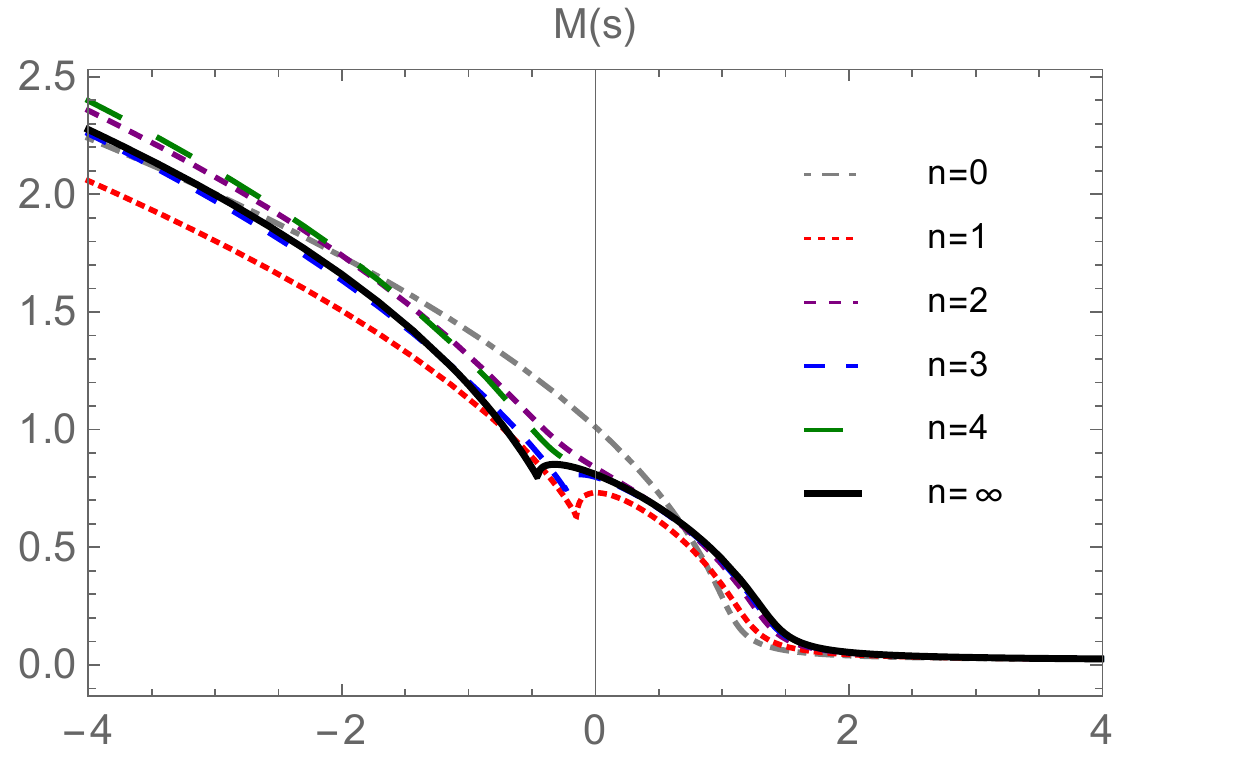}
\caption{\label{fig:a_u}
Dressing functions $A(s)$ and $M(s)$ for different values of $n$, 
corresponding to a quark mass $m_u=0.01$GeV.}
\end{figure*}

The situation becomes clearer, when we employ the effective interaction from Eq.~(\ref{eq:mnmodel}) to get
the algebraic form
\begin{eqnarray}\nonumber 
\Gamma^M(P;k) &=& -\frac{1}{2}\left[ \gamma_\mu \chi^M(P;k)\,\Gamma_\nu^\mathcal{C}(k_-)  \right.\\ \nonumber
&+&  \gamma_\mu  S(k_+)  \, \Lambda^M_\nu(P;k) +  \Gamma_\mu^\mathcal{C}(k_+)\chi^M(P;k)\gamma_\nu\\
&+& \left. \Lambda^M_\mu(P;k)  S(k_-)  \gamma_\nu  \right]\;, \label{eq:mngenbsekernel}
\end{eqnarray} 
where we have written $\Lambda^M_\nu(P;k):=\Lambda^M_\nu(P;k;k)$ and denote the dependence 
on the parameter $\mathcal{C}$ explicitly.

One thus finds \cite{Bender:2002as,Bhagwat:2004hn}:
\begin{eqnarray} \nonumber 
\frac{1}{\mathcal{C}}\Lambda_{\nu,n}^M(P;k) &= & -\gamma_\rho \chi^M(P;k)\Gamma_{\nu,n-1}^\mathcal{C}(k_-)S(k_-) \gamma_\rho\\ \nonumber
&-& \gamma_\rho S(k_+) \Gamma_{\nu,n-1}^\mathcal{C}(k_+) \chi^M(P;k) \gamma_\rho \\
&-& \gamma_\rho S(k_+) \Lambda_{\nu,n-1}^M(P;k) S(k_-) \gamma_\rho\;, \label{eq:recbiglambda}  
\end{eqnarray} 
where one has to take care of attributing the correct quark flavors and factors of $S$ and $\Gamma_\nu$
according to their arguments with respect to the subscripts ${}_\pm$, as described above.
This expression can be computed, the recursive terms for $\Lambda_{\nu,i}^M(P;k)$ summed up
to any desired order, and the resulting algebraic equations can be solved. It is helpful
to note here that also for $n\rightarrow\infty$ the equations can be solved explicitly
via geometric summation.

The initial condition for this recursion relation is \cite{Bender:2002as}
\begin{equation}
\Lambda_{\nu,0}^M(P;k) = 0 \;,
\end{equation} 
which (only) in the pseudoscalar case for equal-mass quarks and $\eta=1/2$ implies \cite{Bender:2002as}
\begin{equation}
\Lambda_{\nu,0}^{PS}(P;k) = 0 \quad\Rightarrow\quad \Lambda_{\nu}^{PS}(P;k) \equiv 0\;,
\end{equation} 
which is an excellent testing case for our general setup.
Further details on this construction are rather technical and are thus collected in 
App.~\ref{sec:pskerneldetails}.

\section{Results and Discussion}\label{sec:results}

In the present investigation we ask the question how important corrections to 
the popular RL truncation are in the case of mesons with unequal-mass constituents.
While our results allow not only qualitative, but also quantitative statements,
caveats are due to the simplicity of the interaction and the resulting oversimplification
of structures related to relative quark-antiquark momentum, as well as the resulting 
artificial dependence on the momentum-partitioning parameter $\eta$. As a result, we not
only attempt to study and quantify effects related to the number of recursion steps used
for the QGV and the BSE kernel, which is our main objective, but also investigate
the dependence of relevant quantities on $\eta$. The subsection on the 
quark DSE results, however, is independent of $\eta$.

Our model parameters are fixed to: 
$\mathcal{C}=0.51$ sets the amount and quality
of dressing in the DSE of the QGV, and is fixed throughout this paper, as mentioned above.
Investigations of the effect of using different values of $\mathcal{C}$ have been
performed in \cite{Bhagwat:2004hn} for equal-mass constituents to some extent, but 
mainly concerning the change from the abelian value of $\mathcal{C}=-1/8$ to a
non-abelian-dominated version, which is phenomenologically reasonable and extendable.
A detailed investigation of the $\mathcal{C}$-dependence of our results is beyond
the present study and thus deferred to subsequent publications.

The effective coupling is set to $\mathcal{G}=0.69$, a value compatible and preadjusted
to fit vector-meson masses throughout entire quark-mass range. This fact also explains
why our results for pseudoscalar mesons as they are presented below in Fig.~\ref{fig:expcomp} 
do not appear completely satisfactorily on the level of a pure theory-experiment comparison.
This is, however, not the main point of our study and serves mainly to keep all
masses and quantities under investigation on their respective domains of reasonable values.

Finally, we also take the current-quark masses directly from \cite{Bhagwat:2004hn} and set  
$m_u=0.01$GeV, $m_s=0.166$GeV, $m_c=1.33$GeV, $m_b=4.62$GeV. For the pion we assume
isospin symmetry and the equality of the current-quark masses of the $u$ and $d$ quarks.

\begin{figure*}[t]
  \includegraphics[width=0.9\textwidth]{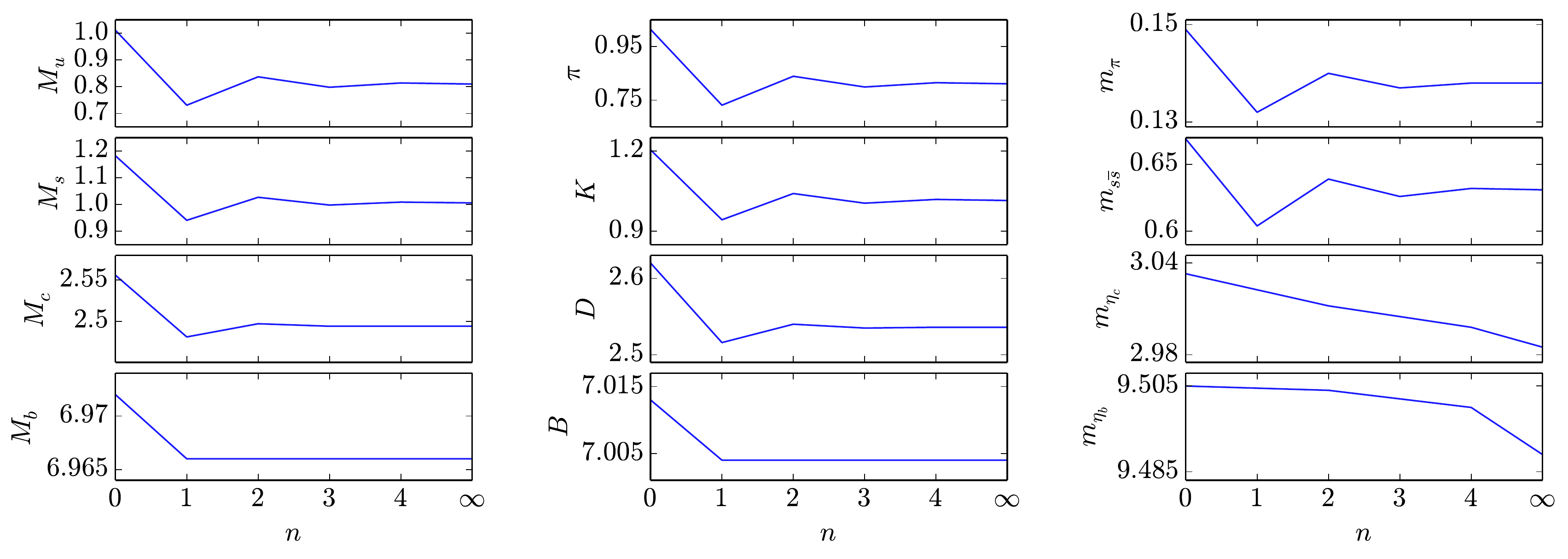}
\caption{\label{fig:mfconvergence}
$n$-dependence of various quantities: dressing functions $M(s)$ and meson masses, all in GeV. 
Left panel: $M(s=0)$ for all four quark masses; center panel: Calculated mass function $M(s)$ 
at $s=-m_H^2/4$. For the unequal-mass case the mass function of  the heavier quark is presented. 
The experimental values are used for the meson masses $m_H$ (see text). The labels $H$ of the 
subplots denote $\pi$, $K$, $D$, $B$; right panel: Pseudoscalar bound-state masses for the
equal-mass case calculated from the BSE for different quark flavors.}
\end{figure*}

\begin{figure}[b]
  \includegraphics[width=\columnwidth]{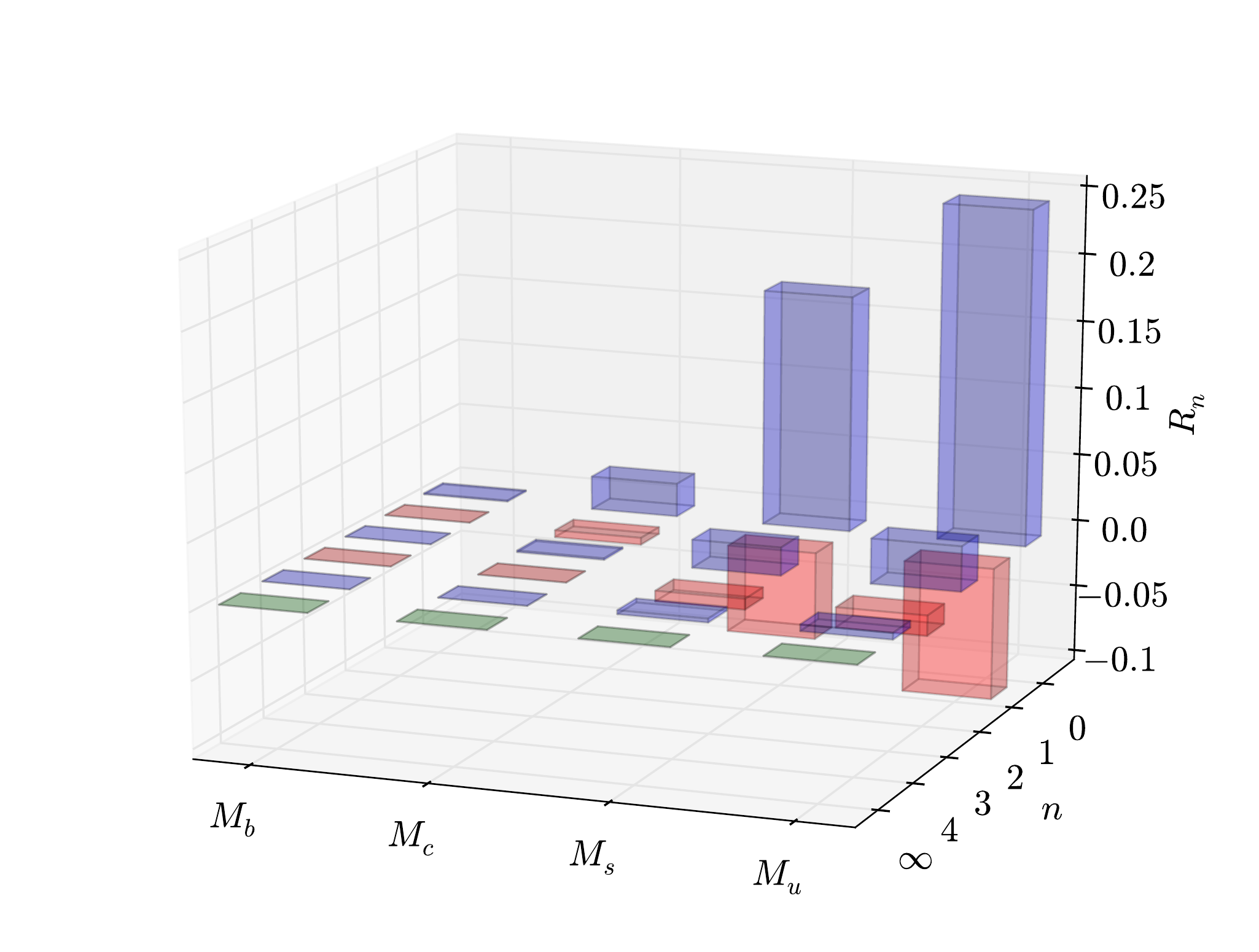}
\caption{\label{fig:mfconvergence3d}
Relative differences $R_n$ of dressing functions $M(0)$ for $n$ loops compared to the case 
$n\rightarrow\infty$, for all four quark masses.}
\end{figure}

\subsection{Quark DSE}

\begin{figure*}[t]
  \includegraphics[width=0.95\columnwidth]{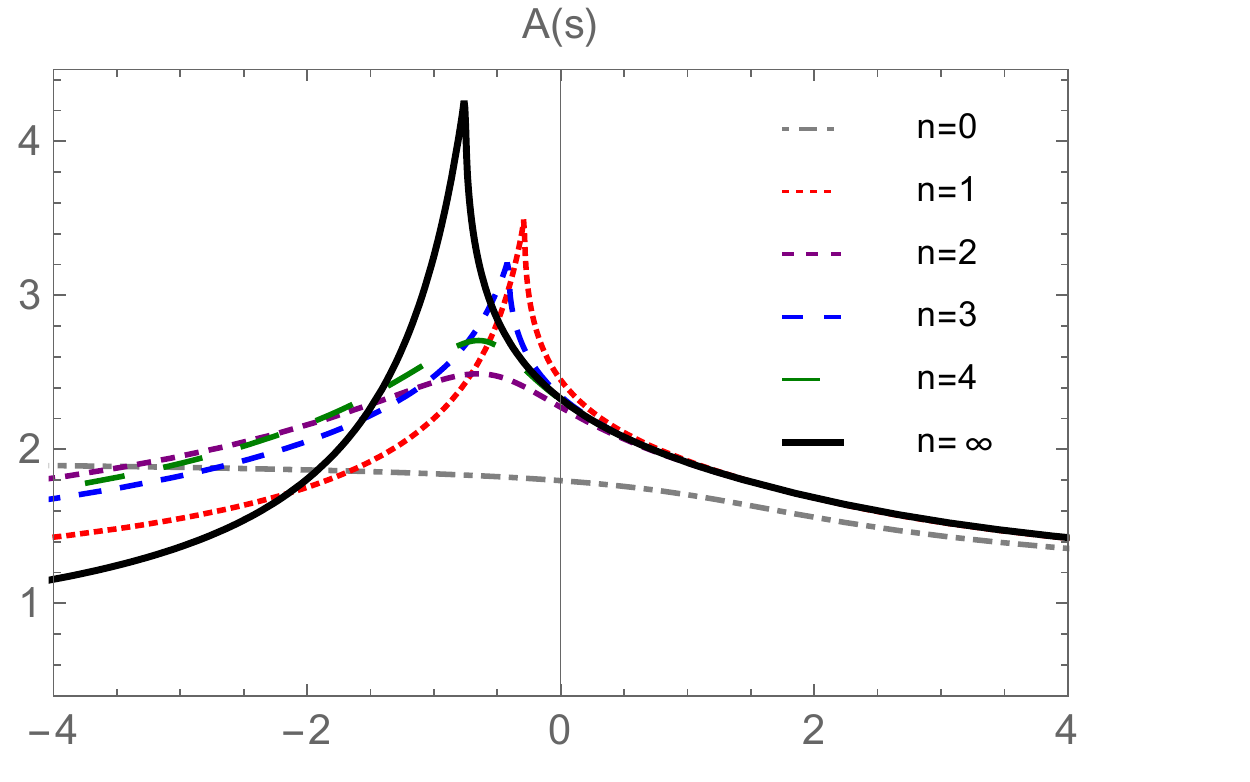}
  \includegraphics[width=0.95\columnwidth]{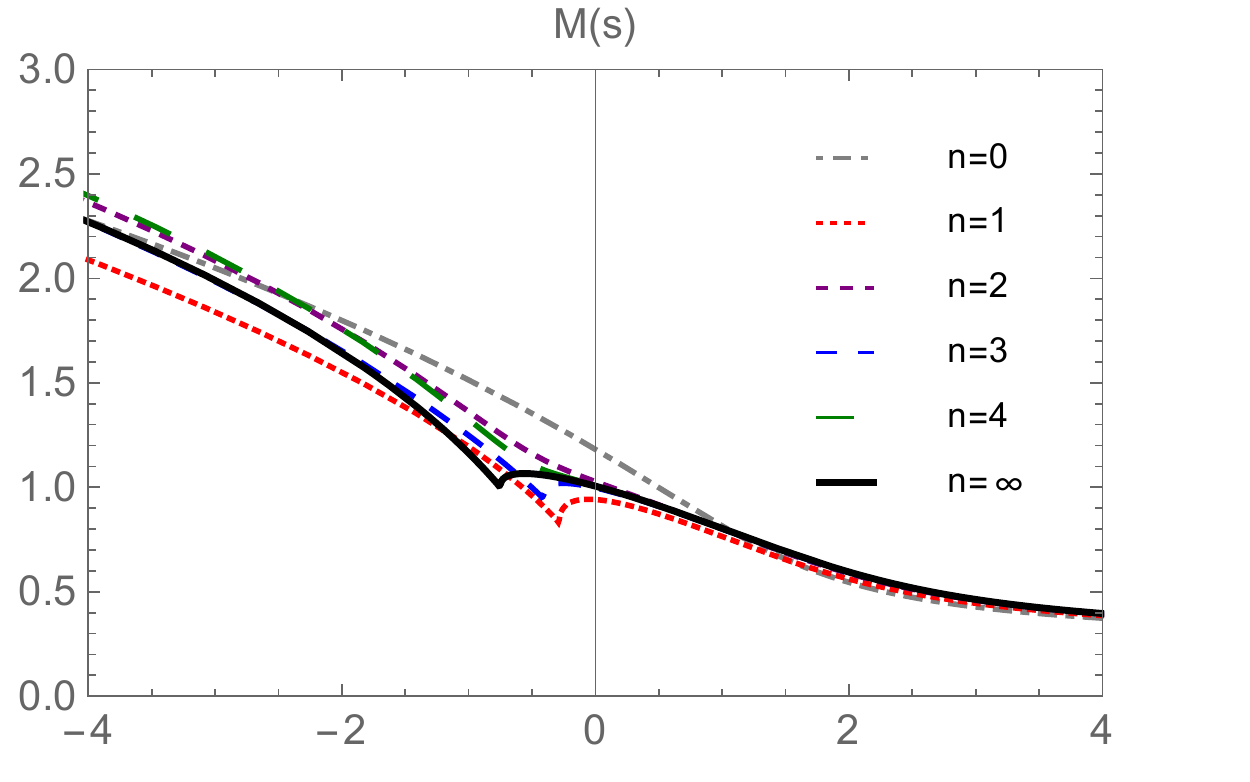}\vspace*{1ex}
  
  \includegraphics[width=0.95\columnwidth]{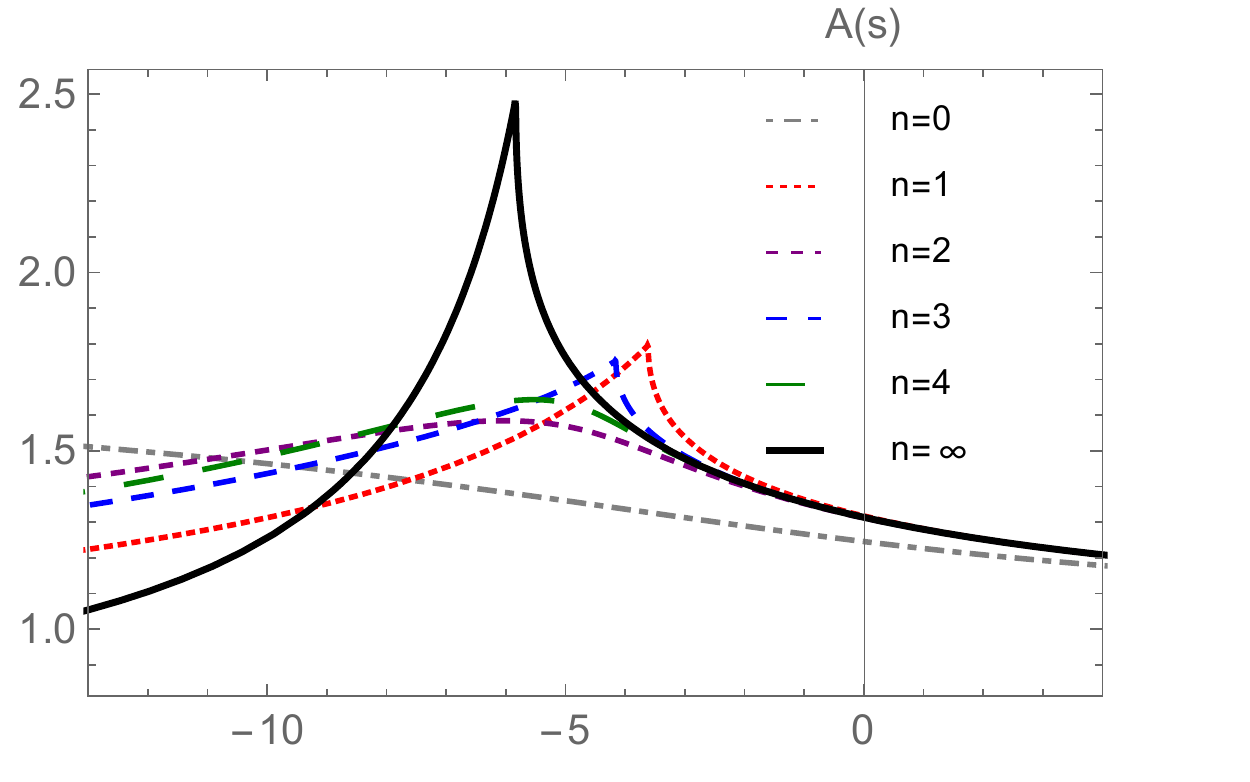}
  \includegraphics[width=0.95\columnwidth]{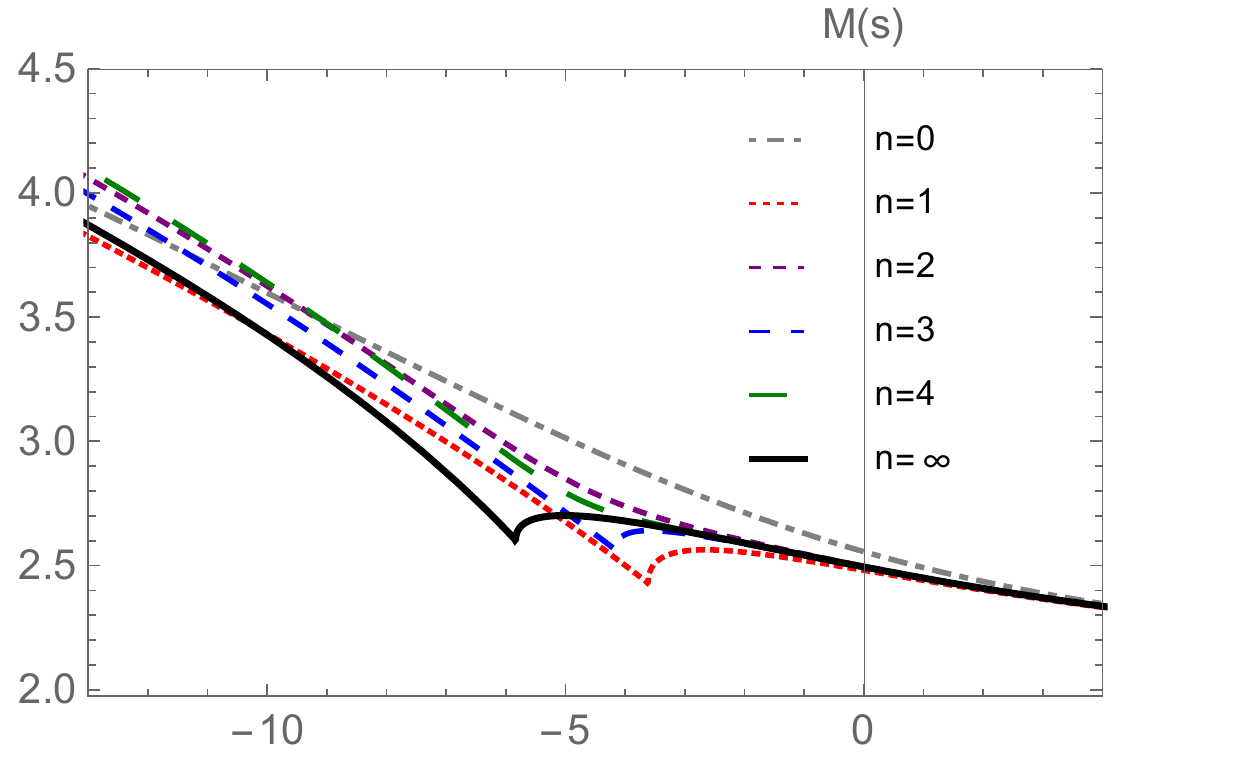}\vspace*{1ex}
  
  \includegraphics[width=0.95\columnwidth]{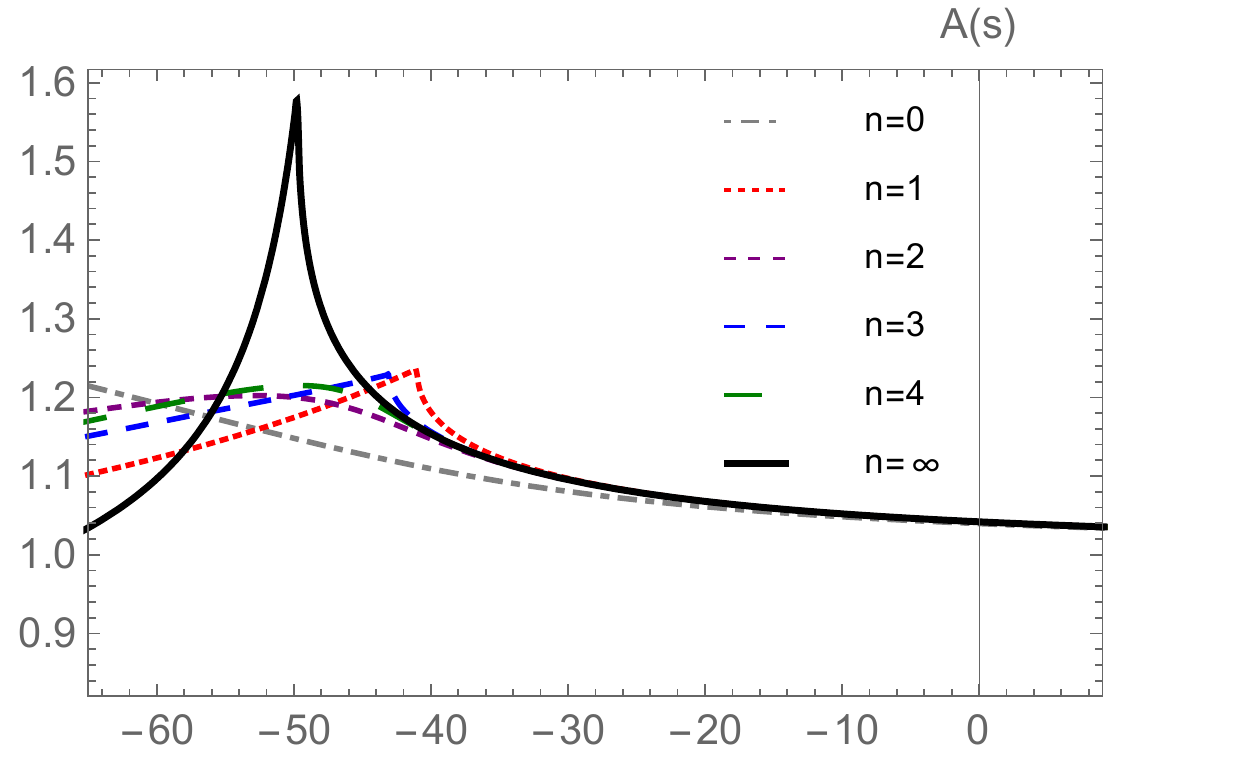}
  \includegraphics[width=0.95\columnwidth]{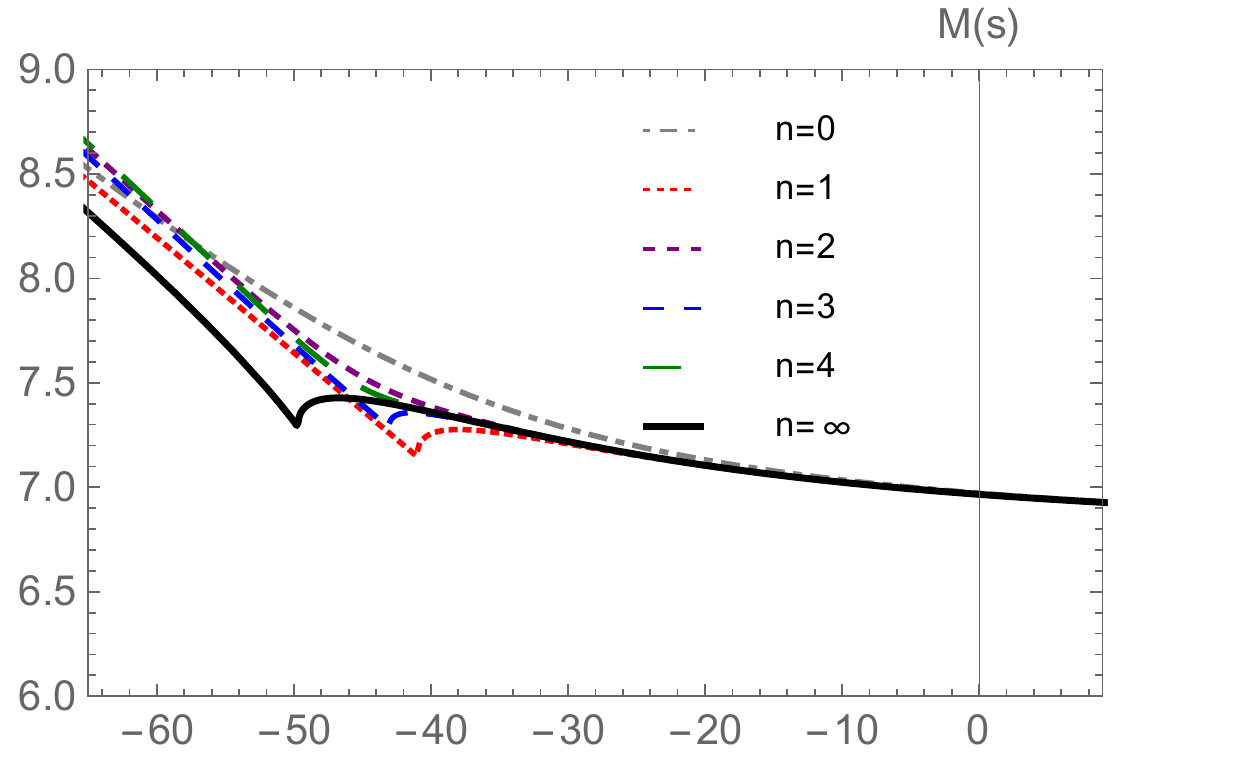}
\caption{\label{fig:amall}
Quark propagator dressing functions $A(s)$ and $M(s)$ as functions of $n$, where
$n=0$ and $n=\infty$ are represented by the dashed and thick solid lines, respectively.
Upper left and right panels: $A(s)$ and $M(s)$ for $m_s=0.166$GeV;
middle left and right panels: $A(s)$ and $M(s)$ for $m_c=1.33$GeV;
lower left and right panels: $A(s)$ and $M(s)$ for $m_b=4.62$GeV.
}
\end{figure*}

The equations resulting from the general setup described in Sec.~\ref{sec:setup}, after
applying the effective interaction defined in Eq.~(\ref{eq:mnmodel}), are algebraic and
can be tackled via standard root-finding techniques. We note here that the solutions
from equations involving polynomials can be defined piecewise, as is also the case in 
our investigation and visible in Figs.~\ref{fig:a_u} and \ref{fig:amall}.

A typical set of solutions for the dressing functions $A(s)$ and $M(s)$ is plotted
in Fig.~\ref{fig:a_u} for the light quark mass and different
\begin{table}[b]
\caption{Calculated mass function $M(s)$ at $s=0$ for light and heavy quarks, in GeV. \label{tab:mofzero}}
\begin{ruledtabular}
\begin{tabular}{ c  c  c  c  c }
               &   $M_u (s=0)$   &  $M_s(s=0)$     & $M_c (s=0)$  &  $M_b (s=0)$      \\ \hline 
   $n=0$       &  1.011   &   1.182 &  2.556      &  6.972             \\ 
   $n=1$       &  0.731   &   0.941 &  2.481      &  6.966            \\
   $n=2$       &  0.837   &   1.027 &  2.497      &  6.966            \\
   $n=3$       &  0.798   &   0.998 &  2.494      &  6.966            \\
   $n=4$       &  0.814   &   1.009 &  2.494      &  6.966            \\ \hline
   $n=\infty$  &  0.810   &   1.006 &  2.494      &  6.966            \\ 
\end{tabular}
\end{ruledtabular}
\end{table}
steps in the recursion, starting from $n=0$, which corresponds to RL truncation, and including 
the fully dressed vertex represented by $n=\infty$. Analogous plots for the 
strange, charm, and bottom masses can be found in Fig.~\ref{fig:amall} in App.~\ref{sec:masssolutions}.

It is notable, how the simple behaviour of $A$ and $M$ are changed from the
RL result (dashed lines) with the QGV dressing switched on. The change is
more drastic for $A$, where one can find qualitatively different trends on 
the timelike domain at the scales relevant for the respective bound states, i.\,e., 
$s=-m_M^2/4$, where $m_M$ is the bound state's mass.
For $M$ the changes are less pronounced; however they are certainly quantitatively 
relevant. In addition, $M$ appears together with $A$ in the denominator of the 
dressed propagators, where their effects are combined.

To illustrate this further, we present both tabulated concrete values as well as plots.
First of all, we tabulate the mass function $M(s)$ at $s=0$, which is one possible definition 
of a constituent-quark mass, in Tab.~\ref{tab:mofzero}. 
The same information is plotted in the left panel of Fig.~\ref{fig:mfconvergence}. One observes
two main features: First, the results seem to alternate in the sign of the effect relative to the 
fully dressed result with respect to $n$: odd $n$ yields a lower result than $n\rightarrow\infty$, 
even $n$ a higher one. Secondly, as expected, dressing effects are weaker in the heavy-quark domain. 
While the changes are pronounced in the light-quark case, a result identical to the one for $n\rightarrow\infty$
is reached already for $n=3$ for the charm quark, and $n=1$ for the bottom quark, respectively.
Defining the relative changes from RL truncation ($n=0$) to the fully dressed result ($n\rightarrow\infty$) as
\begin{equation}\label{eq:reldiff}
R_n:=\frac{M_n(s=0)-M_\infty(s=0)}{M_\infty(s=0)} \;,
\end{equation} 
we obtain $R_0$-values of $0.25$ for the $u$ quark, $0.17$ for the $s$ quark, 
$0.025$ for the $c$ quark, and $0.00086$ for the $b$ quark. This comparison, together with values of $R_n$ for all $n$ and
quark masses are plotted in Fig.~\ref{fig:mfconvergence3d}.

\begin{table}[b]
\caption{Calculated mass function $M(s)$ at $s=-m_H^2/4$, in GeV. For the unequal-mass case the mass function of 
the heavier quark is presented. The experimental values are used for the meson masses $m_H$ (see text).
\label{tab:mofmpi4}}
\begin{ruledtabular}
\begin{tabular}{ c  c  c  c  c  c  c}
    $H$         &   $\pi$       &   $K$         &  $D$         &  $B$       &  $\eta_c$      &  $\eta_b$         \\ \hline 
   $n=0$        &    1.014     	&   1.204       &   2.620      &   7.013    &  2.733         &   7.157       \\ 
   $n=1$        &    0.731    	&   0.943       &   2.516      &   7.004    &  2.559         &   7.123       \\
   $n=2$        &    0.839    	&   1.041       &   2.540      &   7.004    &  2.613         &   7.126       \\
   $n=3$        &    0.799     	&   1.005       &   2.535      &   7.004    &  2.598         &   7.126       \\
   $n=4$        &    0.815    	&   1.019       &   2.536      &   7.004    &  2.603         &   7.126       \\ \hline
   $n=\infty$   &    0.811    	&   1.015       &   2.536      &   7.004    &  2.602         &   7.126       \\ 
\end{tabular}
\end{ruledtabular}
\end{table}
In a similar fashion, we present values for the mass functions for different
quarks at that value of $s$ which appears as the argument of the dressing functions of $S$ 
in the solution of the BSE. For the case of the effective interaction (\ref{eq:mnmodel}) used 
herein this is proportional to $\eta^2$ and the pseudoscalar bound-state mass-squared.
While we investigate the entire range of values of $\eta$ from $0$ to $1$ below, we use $\eta=0.5$ 
for the purpose of this argument and set $s=-m_H^2/4$. The corresponding numbers are collected in 
Tab.~\ref{tab:mofmpi4} and the first four columns are also plotted in the center panel of 
Fig.~\ref{fig:mfconvergence}. For the unequal-quark-mass case we give/plot values of the 
mass function corresponding to the heavier quark flavor. In order to avoid additional uncertainty,
the experimental values are used for the meson masses $m_H$ for each meson as given in the column labels: 
$m_{\pi}=0.140$ GeV, $m_{K}=0.494$ GeV, $m_{D}=1.870$ GeV, $m_{B}=5.279$ GeV, $m_{\eta_c}=2.984$ GeV, 
and $m_{\eta_b}=9.398$ GeV. It should be noted here that in the heavy-light case with values of 
$\eta$ ranging from $0$ to $1$ one potentially covers the whole range $s=-m_H^2$ to $s=0$, corresponding
to $\eta=1$ and $\eta=0$, respectively. This is particularly interesting regarding the behavior
of the dressing functions with $n$ as discussed in more detail in App.~\ref{sec:masssolutions}

The pattern in Tabs.~\ref{tab:mofzero} and \ref{tab:mofmpi4} are similar, which confirms a consistent 
picture of the dressing effects on the functions $A(s)$ and $M(s)$ on the entire relevant parts of the 
timelike $s$ domain. This remark remains also valid if one uses a more sophisticated effective interaction
than the one of Eq.~(\ref{eq:mnmodel}): in such a case, the integral-equation character of the BSE remains
intact and the domain on which $A(s)$ and $M(s)$ are sampled most prominently in the complex $s$ plane
is the one surrounding the negative real axis, reaching out to a scale of $s=-m_H^2/4$; for more
details, see the appendices of \cite{Krassnigg:2010mh,Blank:2011ha,Dorkin:2013rsa,Hilger:2014nma,Dorkin:2014lxa}.

It is illustrative to note here that the alternating convergence pattern between even and odd $n$ can be
attributed to the negative sign of the r.h.s.\ of Eq.~(\ref{eq:qgvrecursion}), the recursion relation for the QGV.

\subsection{ BSE}

\begin{table}[b]
\caption{Pseudoscalar bound-state masses for equal quark masses calculated from the BSE for different
quark masses, in GeV. \label{tab:pseqmass}}
\begin{ruledtabular}
\begin{tabular}{ c  c  c  c  c }
           &  $\pi$ &   $0^{-+}_{\bar{s}s}$  &  $\eta_{c}$  &  $\eta_{b}$      \\ \hline  
   $n=0$     					&   0.149    &  0.669   &  3.033    &  9.505        \\
   $n=1$                        &   0.132    &  0.604   &  ...      &  ...          \\
   $n=2$                        &   0.140    &  0.639   &  3.012    &  9.504        \\
   $n=3$                        &   0.137    &  0.626   &  ...      &  ...          \\
   $n=4$                        &   0.138    &  0.632   &  2.998    &  9.500       \\  \hline
   $n=\infty$                   &   0.138    &  0.631   &  2.985    &   9.489       \\ 
\end{tabular}
\end{ruledtabular}
\end{table}

With the details of the solutions of the quark DSE laid out and the solutions obtained for the four
relevant quark flavors, we proceed to the corresponding results of the pseudoscalar meson BSE.
As a first test and for easy comparison and anchoring, we set $\eta=1/2$ and recompute the 
equal-quark-mass results of Ref.~\cite{Bhagwat:2004hn}, which we have collected in Tab.~\ref{tab:pseqmass}
and plotted in the right panel of Fig.~\ref{fig:mfconvergence}.
While the emerging convergence pattern is qualitatively similar to the ones observed above for the quark mass
function, we also encounter cases where the BSE does not yield a solution for particular values of $n$, e.\,g.,
$n=1,3$ for the $\eta_{c}$ and $\eta_{b}$ masses. 

\begin{figure*}[t]
  \includegraphics[width=0.9\textwidth, clip=true]{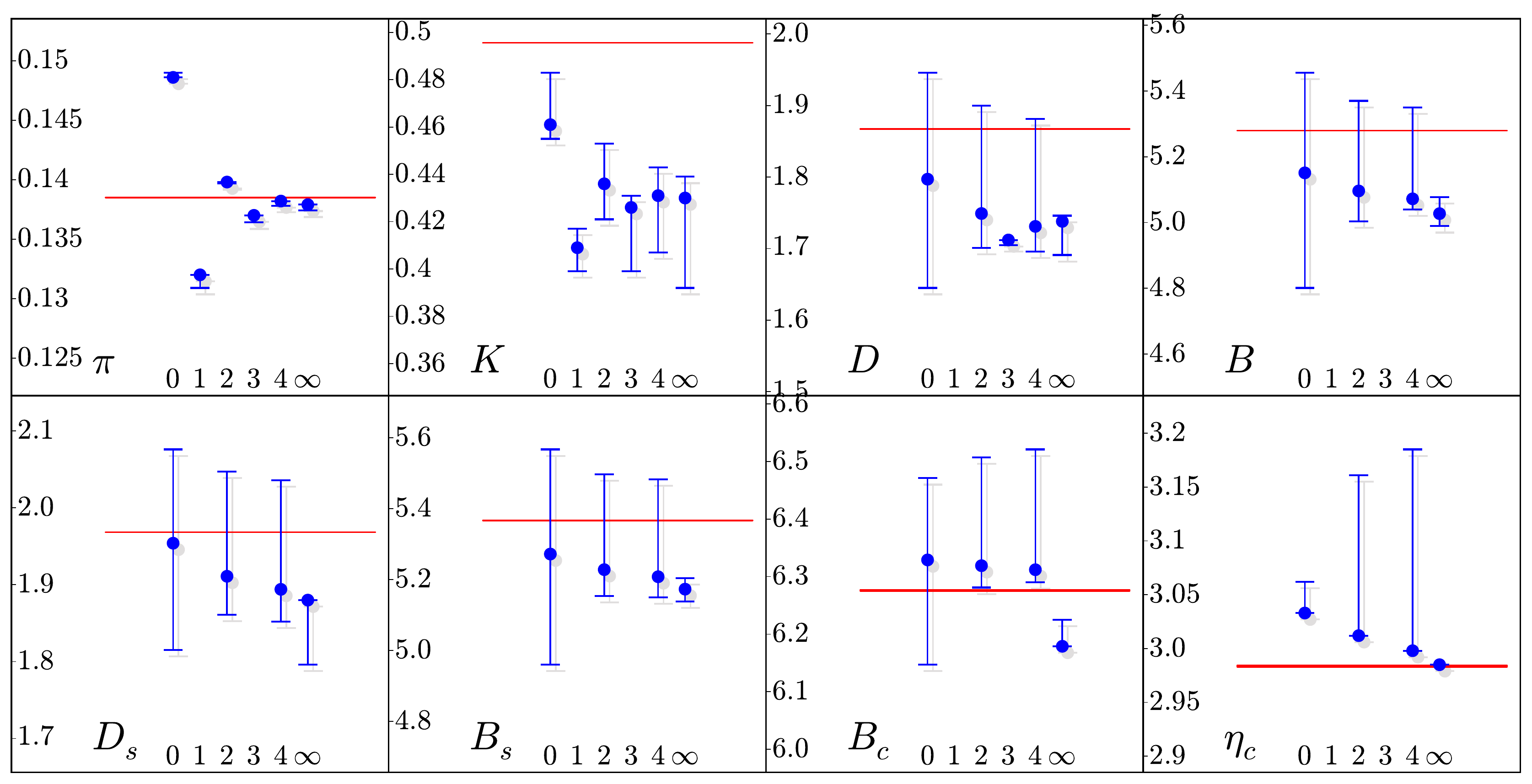}
\caption{\label{fig:expcomp}
Bound-state masses for $\pi$ meson, $\eta_c$, and flavored pseudoscalar ground states as a function of $n$, given in GeV. 
The dependence on $\eta$ is illustrated via the error bars. Calculated results are given by blue dots; experimental
data are represented by horizontal lines.}
\end{figure*}

To understand this, a few remarks are in order. First of all, the homogeneous BSE is an equation, which in principle
does not need to have a solution in all possible cases or for all given configurations in a certain situation or
setup. Depending on truncation setup, model features or defects and also ranges of parameters, one can encounter
cases where there is no solution. In such a case, we denote the missing solution by three dots in the corresponding
spot in the table and curves in figures go across such a point. In Fig.~\ref{fig:expcomp}, for missing results in
a particular case, no calculated data point is generated.

\begin{table}[t]
\caption{Bound-state mass for the kaon as a function of $n$ and $\eta$, given in GeV. \label{tab:mkaon}}
\begin{ruledtabular}
  \begin{tabular}{ c  c  c  c  c  c  c  c }
			    $\eta$   &  $0$    &  $0.2$   &  $0.4$     &  $0.5 $    &   $0.6$  & $0.8$     &  1        \\ \hline
                            $n=0$    &  0.455  &  0.455   &  0.458     &  0.461     &   0.464  &  0.472    &  0.483    \\
                            $n=1$    &  ...    &   ...    &  0.399     &  0.409     &   0.416  &  0.417    &   ...     \\
                            $n=2$    &  0.421  &  0.426   &  0.433     &  0.436     &   0.440  &  0.447    &  0.453    \\
                            $n=3$    &  ...    &  0.399   &  0.420     &  0.426     &   0.431  &  0.431    &  0.427    \\  
                            $n=4$    &  0.407  &  0.416   &  0.426     &  0.431     &   0.435  &  0.441    &  0.443    \\ \hline
  		 $n=\infty$          &  0.392  &  0.411   &  0.425     &  0.430     &   0.434  &  0.439    &  0.438    \\
  \end{tabular}
\end{ruledtabular}
\end{table}

\begin{figure}[b]
  \includegraphics[width=\columnwidth, clip=true]{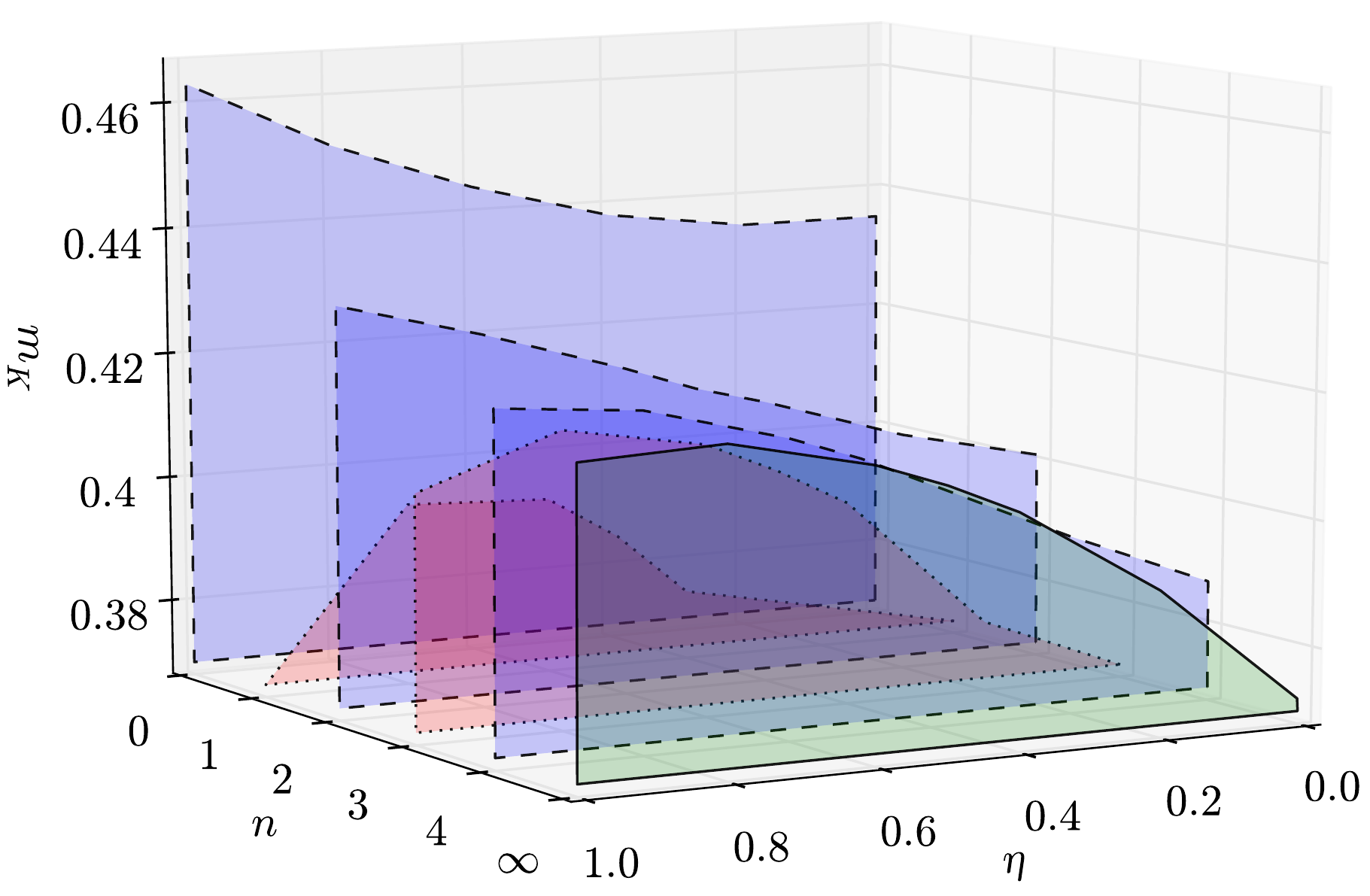}
\caption{\label{fig:mkaon}
Bound-state mass for the kaon as a function of $n$ and $\eta$, given in GeV. Even $n$ are depicted by dashed
lines, odd ones by dotted lines, and the fully summed result by a solid line.}
\end{figure}
That said, we proceed to the case of unequal (anti)quark masses in the meson and present results for the various 
combinations of quark flavors. We start with the kaon, for which the results are presented
in Tab.~\ref{tab:mkaon}. At this point a comment about the $\eta$ dependence of the results
is necessary. Eq.~(\ref{eq:mnmodel}) leads to the model artifact that the relative momentum in all parts of the 
calculation of the meson mass vanishes. This not just simplifies the structure of the equations but also the structures of
the pseudoscalar (and other) BSAs, the QGV, and the correction terms $\Lambda$. One of the results of these 
simplifications is the artificial $\eta$ dependence of the results already mentioned above. To remain in control
of our results and the corresponding discussion, we thus have to analyze and quantify the $\eta$ dependence
of the observables under consideration. We do this by providing a complete set of results as well as encoding
the $\eta$ dependence in the form of a systematic error in our results in Fig.~\ref{fig:expcomp}: the error bars
are plotted from the lowest to the largest value of the mass result for a given $n$. As a result, they are
asymmetric and the value of $\eta$ chosen for the data point, as defined below, may on occasion be also either
the smallest or the largest value available at this $n$.

To illustrate the extent of this effect, we present a concrete example, namely the kaon mass as a function of 
both $n$ and $\eta$ tabulated in Tab.~\ref{tab:mkaon}. As a further illustration, we present the same numbers 
plotted in Fig.~\ref{fig:mkaon}. One observes: convergence with $n$; an alternating convergence pattern with
respect to odd and even $n$ analogous to the ones observed above; and a moderate dependence on $\eta$ in the 
sense that the effect across the entire $\eta$ range is comparable in relative size to the effect of the dressing
from RL truncation to the fully summed result for a fixed value of $\eta$. Concretely, we observe changes of
the order of 10\%, which is in agreement with the analysis in \cite{Munczek:1983dx}, where the authors quote
changes smaller than 15\%. Judging from the particular behavior of the curves in Fig.~\ref{fig:mkaon} at the 
outer boundary of the $\eta\in [0,1]$ interval, one can see that these extreme values of $\eta$ also seem 
to lead to correspondingly extreme values of $m_K$. While it is certainly correct to state that the error bars in 
Fig.~\ref{fig:expcomp} should represent the entire range of $\eta$, it is also fair to remark that in practice
the extreme $\eta$ values may not be representative to an amount that actually justifies the size of these
error bars and we in general regard them as overestimates of more suitably defined systematic errors.

Tables and figures analogous to Tab.~\ref{tab:mkaon} and Fig.~\ref{fig:mkaon} are collected---for all mesons 
considered here and whose masses are presented below in Fig.~\ref{fig:expcomp}---in App.~\ref{sec:etadependence}
in Tab.~\ref{tab:mall} and Fig.~\ref{fig:mall}. There, we have several rows completely filled with dots
indicating that no solutions exist for a given $n$. However, there may still be small domains in between
the values of $\eta$ listed in the figures. While we did not include those in Tab.~\ref{tab:mall} and 
Fig.~\ref{fig:mall} to remain at a comprehensible set of results, we have done a finer search and included them
in the results presented in Fig.~\ref{fig:expcomp}. These cases are easily recognized by their small error bars, 
which we chose not to rescale or blow up artificially; a prominent example is the $n\rightarrow\infty$ value
for the $\eta_c$ mass in the lower right corner of the figure.

Finally, in Fig.~\ref{fig:expcomp} we collect all our results in a compact fashion and compare them to
experimental data. As already mentioned above, the aspect of comparison to experiment is not central in our
argumentation and serves merely to put our results into perspective. What we focus on here are the effects
of the dressing introduced in the QGV for the unequal-mass case as compared to the analysis already
provided in \cite{Bhagwat:2004hn}. In Fig.~\ref{fig:expcomp} we present pseudoscalar ground-state masses
for the pion as a reference and all flavored pseudoscalar mesons up to and including $b$ quarks. The 
respective experimental values are marked by the horizontal lines in each of the subplots, while our calculated
results are given as filled circles with error bars as discussed above. In particular, the filled circles
are obtained via the following $\eta$ values: $0.5$ for the $\pi$, $K$, $\eta_c$, and $\eta_b$, $0.82$ for the $D$, 
$0.91$ for the $B$, $0.8$ for the $D_s$, $0.89$ for the $B_s$, and $0.79$ for the $B_c$.

Clouded to some extent by the systematic errors, we can still observe dressing effects in any given $\eta$ slice 
of our data. Using the values plotted as the filled circles in our figure, we calculate absolute and 
relative changes from the dressing effects in a comparison of the fully dressed result to RL truncation
as follows: the absolute difference in mass $m_H$ for a given meson $H$ is obtained between fully dressed 
and RL result and denoted by $\Delta m_H$. The corresponding results for our set of states depicted in 
Fig.~\ref{fig:expcomp} is tabulated in the first data column of Tab.~\ref{tab:diffscollected} and given in
units of GeV. The second column lists the corresponding relative difference
\begin{table}[t]
\caption{Absolute and relative mass differences for the pion and the unequal-mass pseudoscalar mesons
with all possible flavor combinations. $\Delta m_H$ is given in GeV, the other quantities are dimensionless 
(see text). \label{tab:diffscollected}}
\begin{ruledtabular}
  \begin{tabular}{ l  c  c } 
 $H$ & $\Delta m_H$ & $\Delta m_H^\mathrm{rel}$ \\ \hline 
 $\pi$ &  0.011 & 0.078 \\ 
 $\eta_c$ &  0.048 & 0.016 \\ 
 $\eta_b$ &  0.016 & 0.002 \\ 
 $K$ &  0.031 & 0.072 \\ 
 $D$ &  0.059 & 0.034 \\ 
 $B$ &  0.124 & 0.025 \\ 
 $D_s$ &  0.074 & 0.039 \\ 
 $B_s$ &  0.099 & 0.019 \\ 
 $B_c$ &  0.150 & 0.024 \\ 
  \end{tabular}
\end{ruledtabular}
\end{table}

\begin{equation}\label{eq:reldiffmeson}
\Delta m_H^\mathrm{rel}:=\frac{m_H^{n=0}-m_H^{n\rightarrow\infty}}{m_H^{n\rightarrow\infty}} \;,
\end{equation} 
which is dimensionless and a better point of comparison to, e.\,g., our estimates of the systematic error 
from our model artifacts. While we compare the difference to the fully dressed result here, one may also 
divide by the RL result instead, which can be uniquely related to Eq.~(\ref{eq:reldiffmeson}). Since the 
differences are small, however, the effect of such a change does not affect our discussion. 

The absolute difference is largest in the unequal-mass case involving a $b$ quark, topped by the $B_c$ mass,
which changes by $150$ MeV from RL truncation to the fully-dressed case in sharp contrast to pseudoscalar
bottomonium. While the corresponding relative
differences of $\approx 2$ \% make this effect look small again, one must not forget that this is still much
larger than relevant spectroscopic properties like, e.\,g., the hyperfine splittings between the $0^{-+}$ and
$1^{--}$ ground states. In order to concretely investigate the dressing effects on splittings, we need a
similar result like the one presented here for pseudo scalars in addition also for the vector case. While this
is beyond the scope of the present study, it will be investigated and presented in future publications.

Regarding the relative differences, we observe that for our data points their size is similar to the equal-mass 
case and of the same order of magnitude throughout. The $\eta_c$ and $\eta_b$ relative mass differences
are at the lower end of the set of results presented here. In fact, this is also true if one considers
the absolute mass differences, where the effect tor $\eta_b$ is comparable to the $\pi$ and for $\eta_c$ 
is comparable to the $K$.

It is important to note at this point that the pion results are somewhat influenced by the pion mass' chiral-limit
behavior. In particular, the AVWTI is satisfied at every recursive step $n$ in our setup and thus in each 
case the pion mass is zero. Therefore, the variation of the pion mass with $n$ is expected to be small from 
the very beginning. The typical size of the mass difference of $\approx 10$ MeV confirms this expectation.

Overall, we see that the effects are not negligible, warranting further systematic exploration of effects on
beyond-RL treatment of meson (and baryon) properties in the DSBSE approach. At the same time, our results provide
no reason to disregard careful and comprehensive studies in RL truncation from the outset, which opens the
door for straight-forward and large-scale phenomenological studies in an RL setup.

\section{Conclusions}\label{sec:conclusions}

We have extended a previously and phenomenologically successfully applied systematic DSBSE truncation scheme and setup for
pseudoscalar mesons from the equal to the unequal-mass case for the first time. Constructing and reviewing the
necessary pieces in the pseudoscalar quark-antiquark BSE, we have studied and quantified effects
arising from simplification artifacts in the model used. Still, justification for using
such a simple model comes from the fact that it makes involved investigations of 
truncations and/or other features within the DSBSE approach feasible. Our results are presented
in a comprehensive manner in both tables and figures for easy comparability and straight-forward discussion.
Under the assumption that the $\eta$ dependence apparent in our results can be quantified and 
treated as a systematic error, we have presented meson masses for RL truncation as well as for several steps
in the truncation scheme, including the fully summed case.

The results show clear and understandable patterns and confirm important and central expectations according
to the symmetries at both the chiral and the heavy-quark ends of the quark-mass range available from experimental
data. Dressing effects compared to RL truncation are sizeable, but not overwhelming, which provides support
for both RL studies targeted at phenomenology as well as investigations of hadron properties beyond RL truncation.

Our tests of the validity of an effective \emph{Ansatz} for the quark propagator using constant dressing functions
adds to the consensus about the bottom quark being essentially heavy, while the charm quark is not.

\begin{acknowledgments}
We acknowledge helpful conversations with C.~Popovici, H.~Sanchis-Alepuz, P.\,C.~Tandy, and R.~Williams.
This work was supported by the Austrian Science Fund (FWF) under project no.\ P25121-N27.
\end{acknowledgments}

\appendix

\section{Quark propagator dressing functions}\label{sec:masssolutions}
In this appendix we present figures of the dressing functions $A(s)$ and $M(s)$
for the $s$, $c$, and $b$ quarks, analogous to Fig.~\ref{fig:a_u} presented in 
the main text. This serves to provide a complete set of results together with
the corresponding illustrations. All dimensioned quantities are given in GeV.

\begin{figure*}[t]
  \includegraphics[width=\columnwidth]{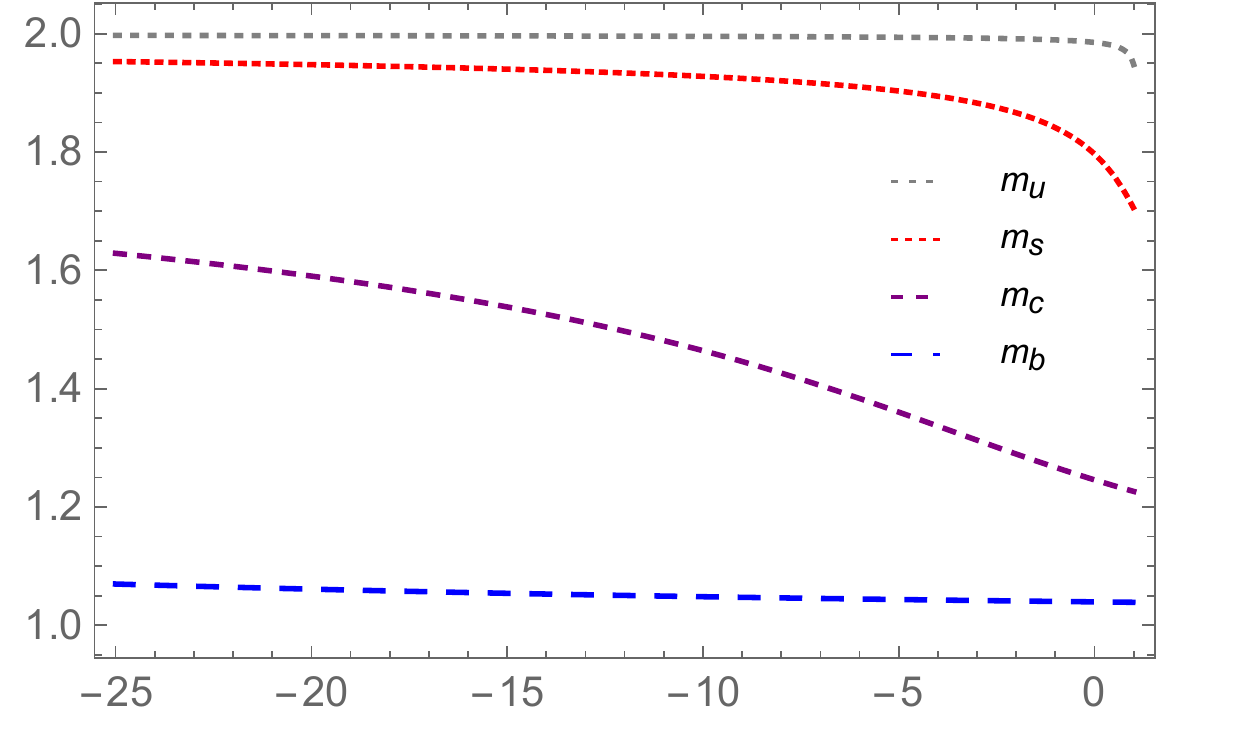}
  \includegraphics[width=\columnwidth]{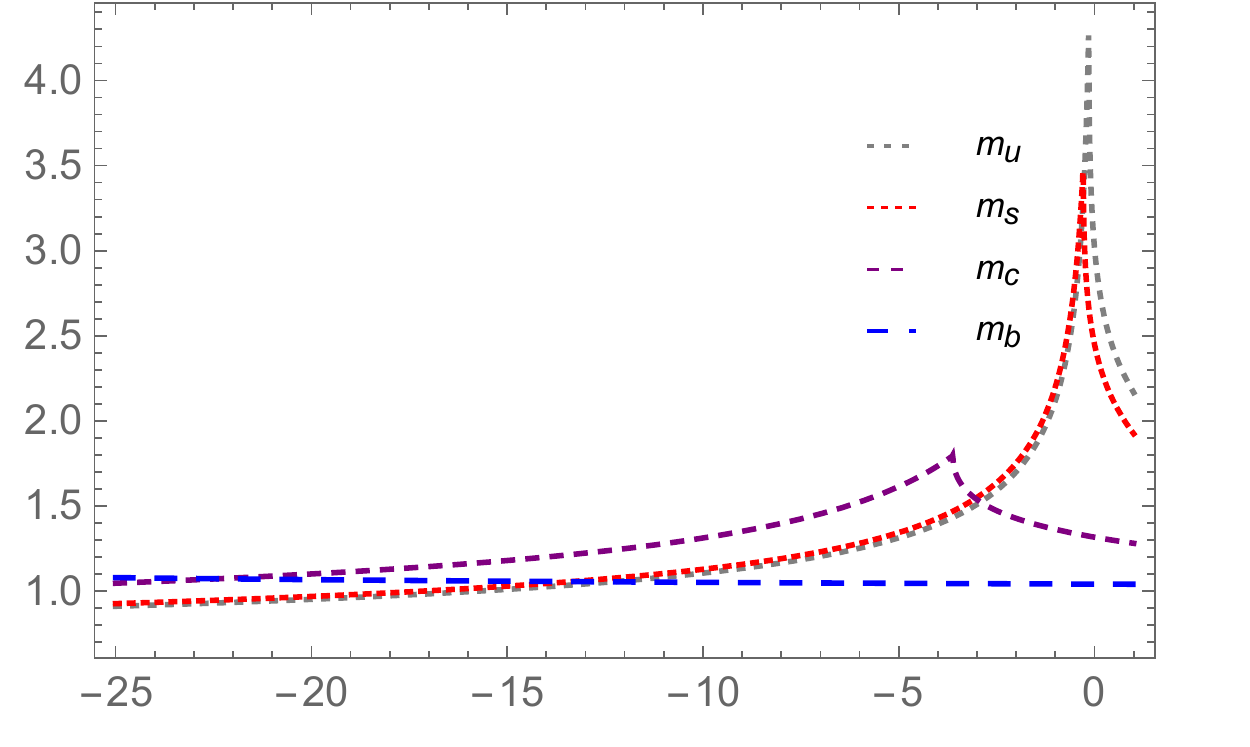}\vspace*{1ex}
  
  \includegraphics[width=\columnwidth]{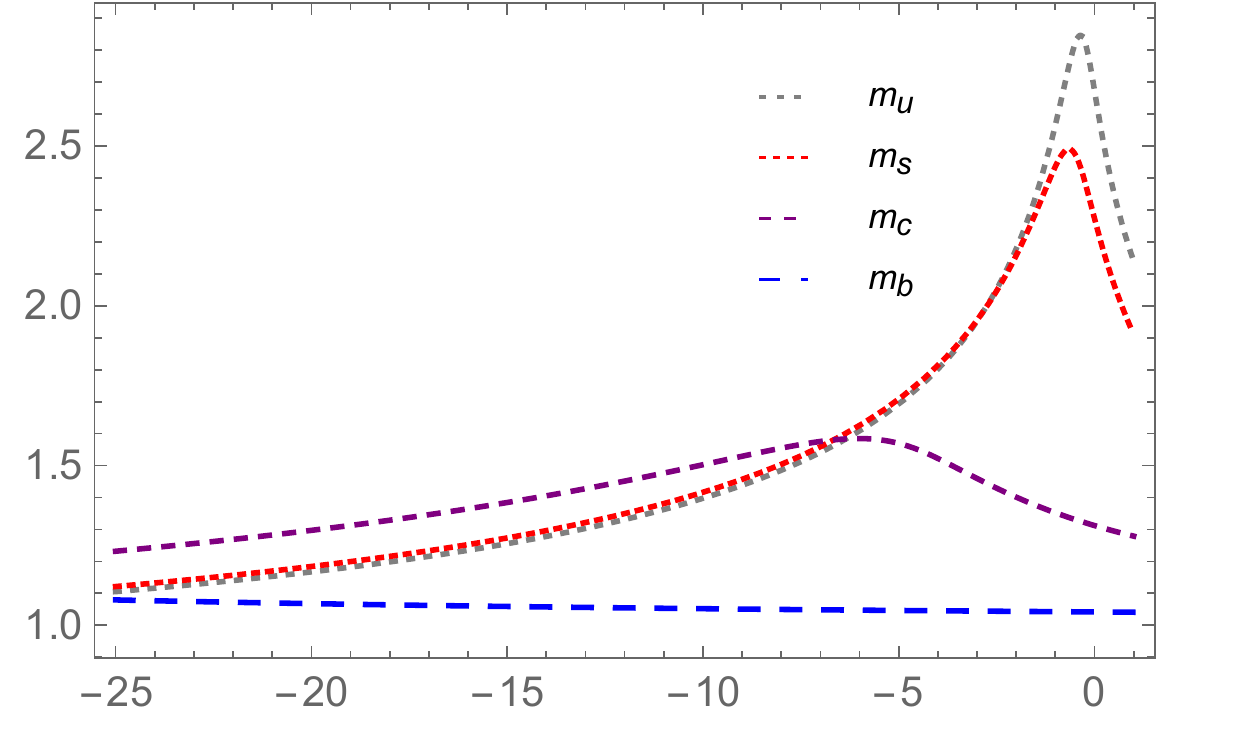}
  \includegraphics[width=\columnwidth]{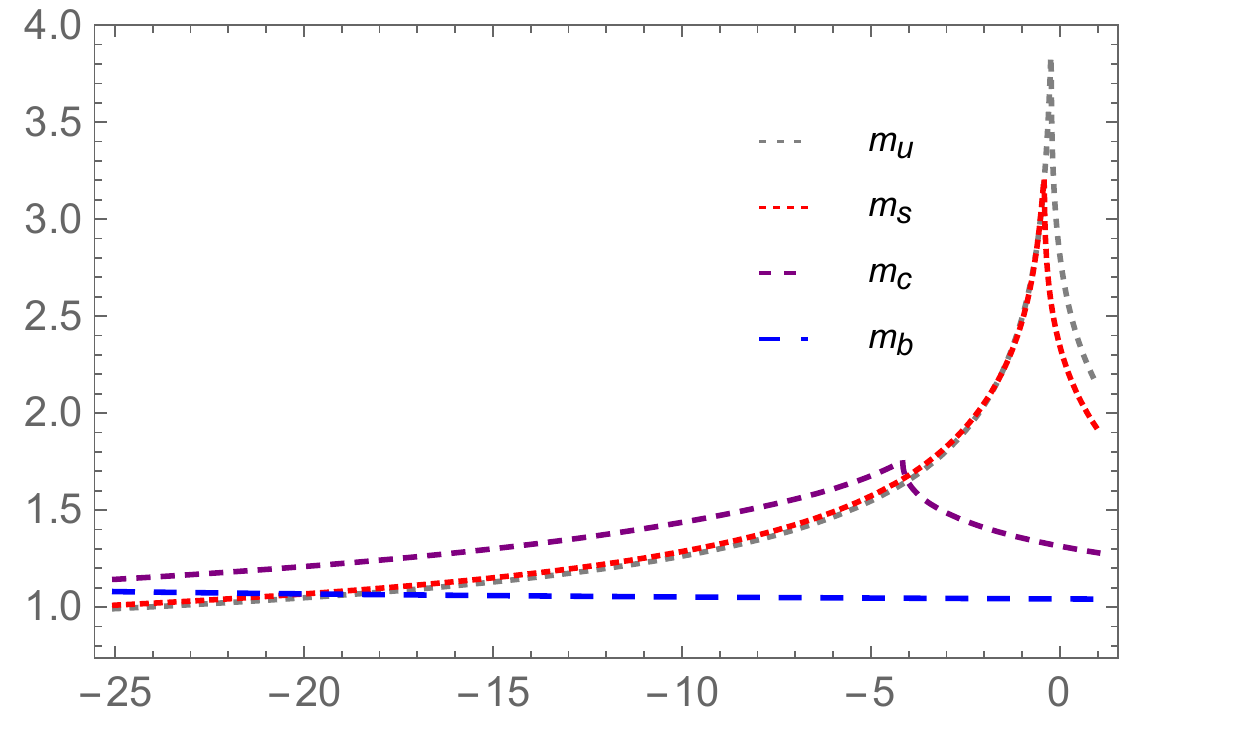}\vspace*{1ex}
  
  \includegraphics[width=\columnwidth]{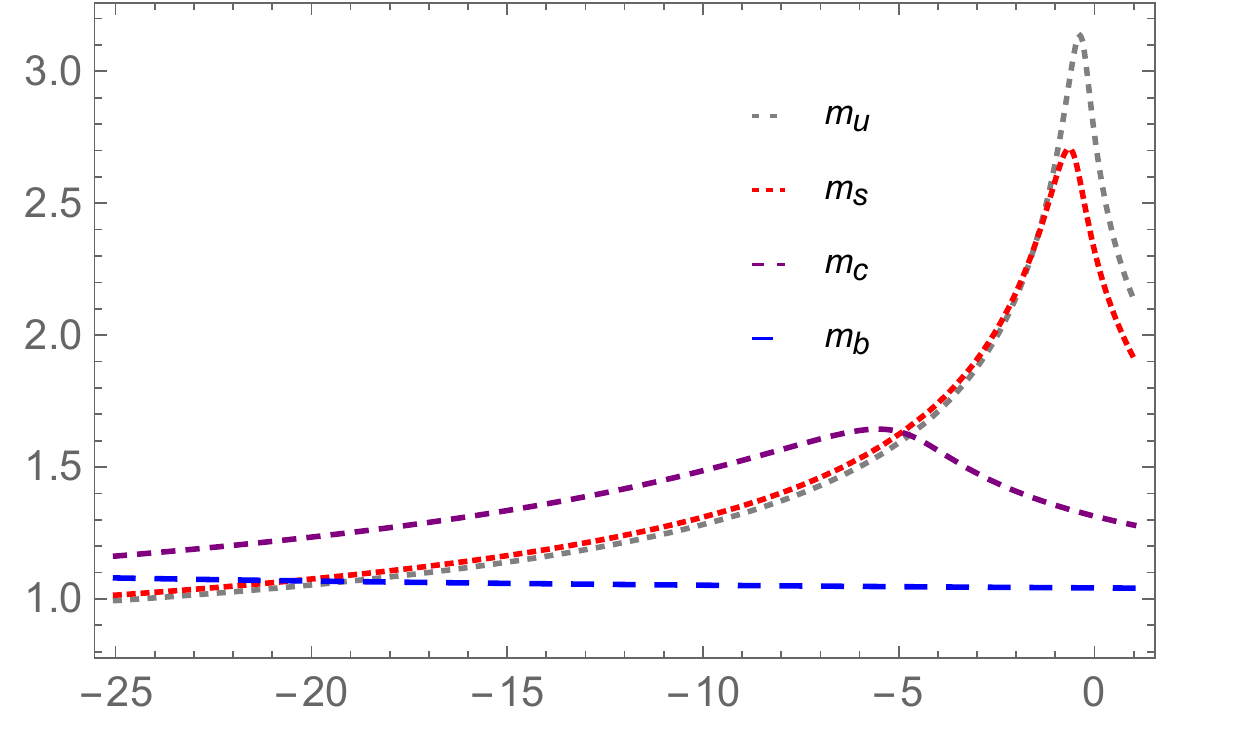}
  \includegraphics[width=\columnwidth]{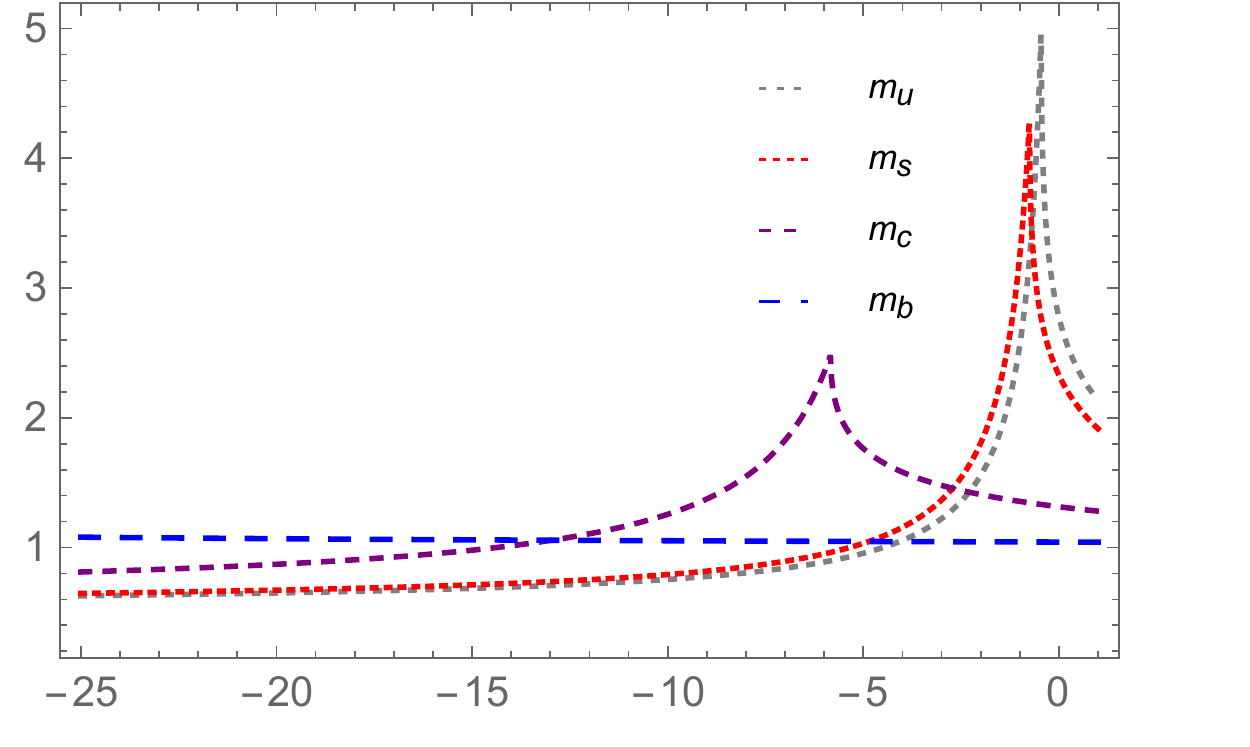}
\caption{\label{fig:adiff}
Comparison of $A(s)$ for the four quark masses, plotted on the same domain for each $n$. Left upper panel: $n=0$; right upper panel: $n=1$;
Left middle panel: $n=2$; right middle panel: $n=3$; Left lower panel: $n=4$; right lower panel: $n=\infty$.}
\end{figure*}

One observes easily how the distinctive features of the curves move further into
the timelike domain with increasing quark mass. In particular, a scale relevant
to the computation of meson properties via the BSE in our setup is the value
$s=-m_H^2/4$ for which $M(s)$ is also tabulated in Tab.~\ref{tab:mofmpi4}.
While for the light mesons the meson mass range up to $2$ GeV implies values
of $s\approx-1$ and higher as the region of interest, the corresponding range for 
bottomonium would be $s\approx -25$ and higher. Apparently, said features in $A$ and
$M$ as well as the regions where the differences among the curves resulting
from the various values of $n$ are most pronounced are relevant regarding the BSE
for light quarks and become unimportant for heavier quarks. To elucidate this further,
we provide a detailed comparison to effective free model-quark propagators 
in App.~\ref{sec:heavyquarkprop}.

At this point it is also interesting to note that, as already indicated in Sec.~\ref{sec:results},
depending on the value of $\eta$ one potentially covers the whole range of arguments $s=-m_H^2$ (for $\eta=1$) to $s=0$ 
(for $\eta=0$) in the quark propagators as they appear in the BSE. In practice this means that one can attempt to choose
$\eta$ such that the distinct structures in the respective dressing functions are avoided by
the interplay of $s_+=-\eta^2 m_H^2$ vs.~$s_-=-(1-\eta)^2 m_H^2$, thus, e.\,g., minimizing
dressing effects in the resulting meson properties.

\section{Heavy-quark propagator}\label{sec:heavyquarkprop}

\begin{figure*}[t]
  \includegraphics[width=\columnwidth]{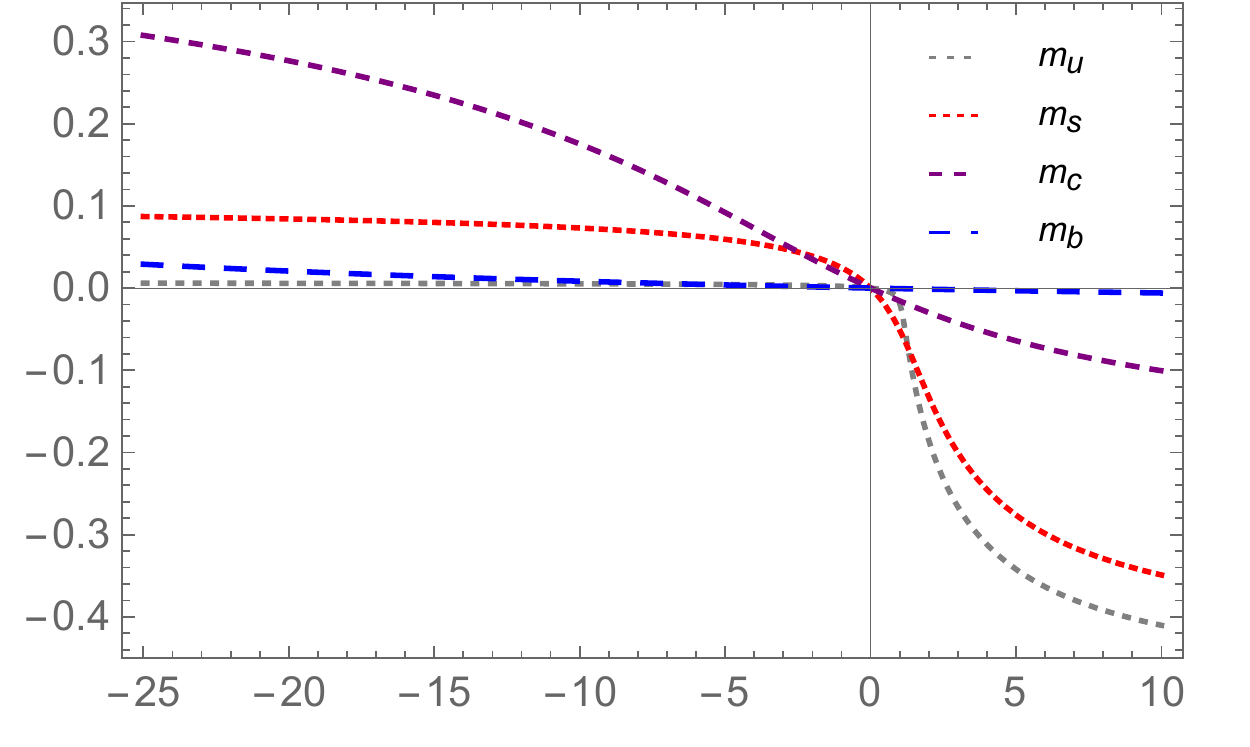}
  \includegraphics[width=\columnwidth]{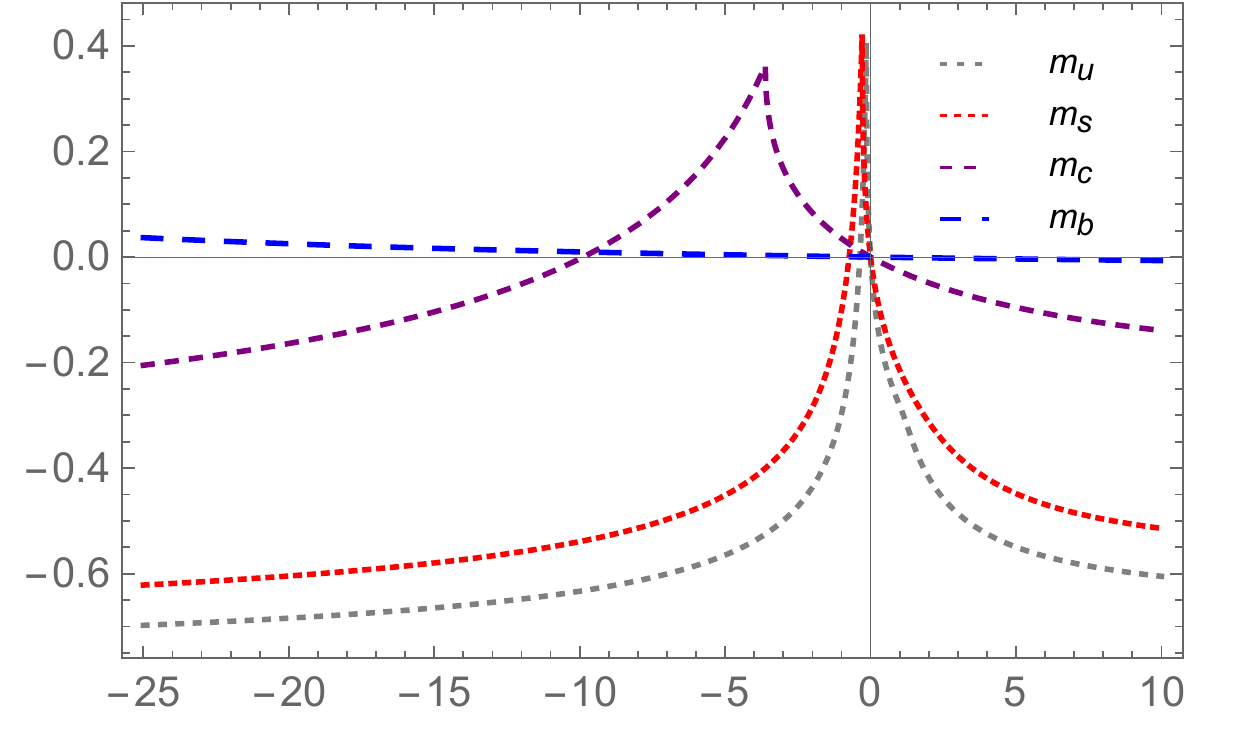}\vspace*{1ex}
  
  \includegraphics[width=\columnwidth]{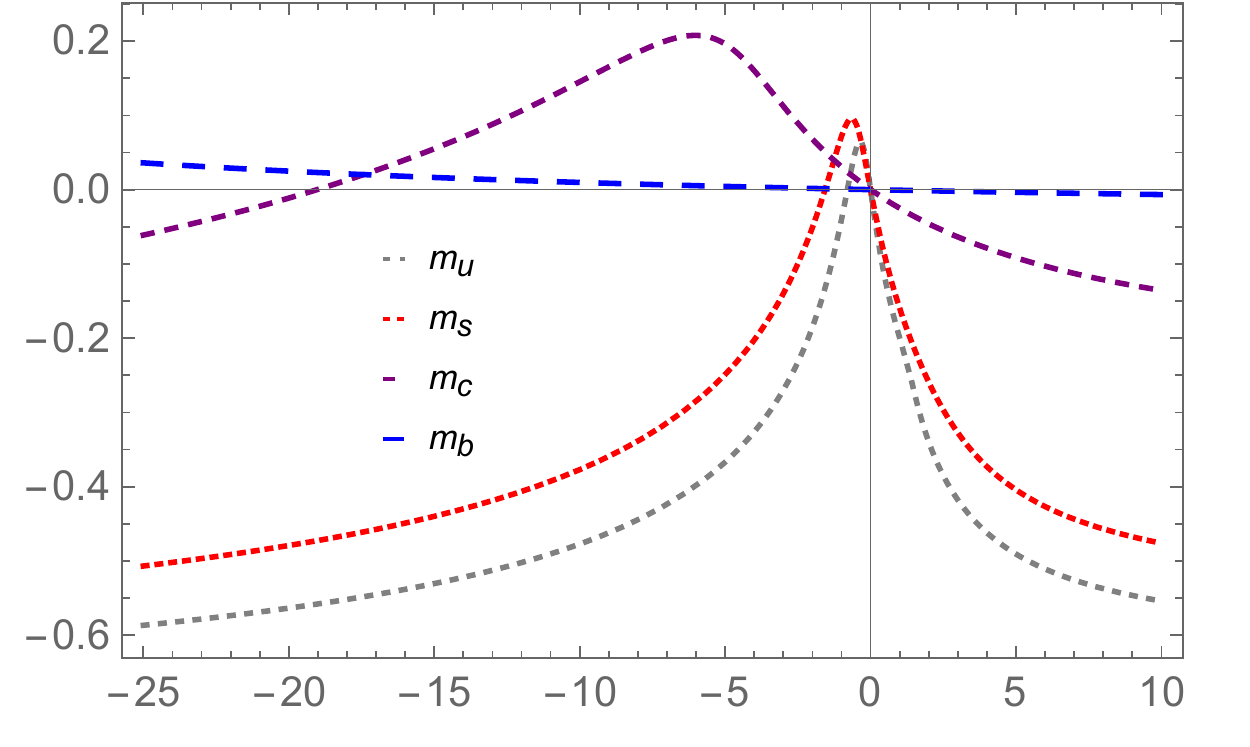}
  \includegraphics[width=\columnwidth]{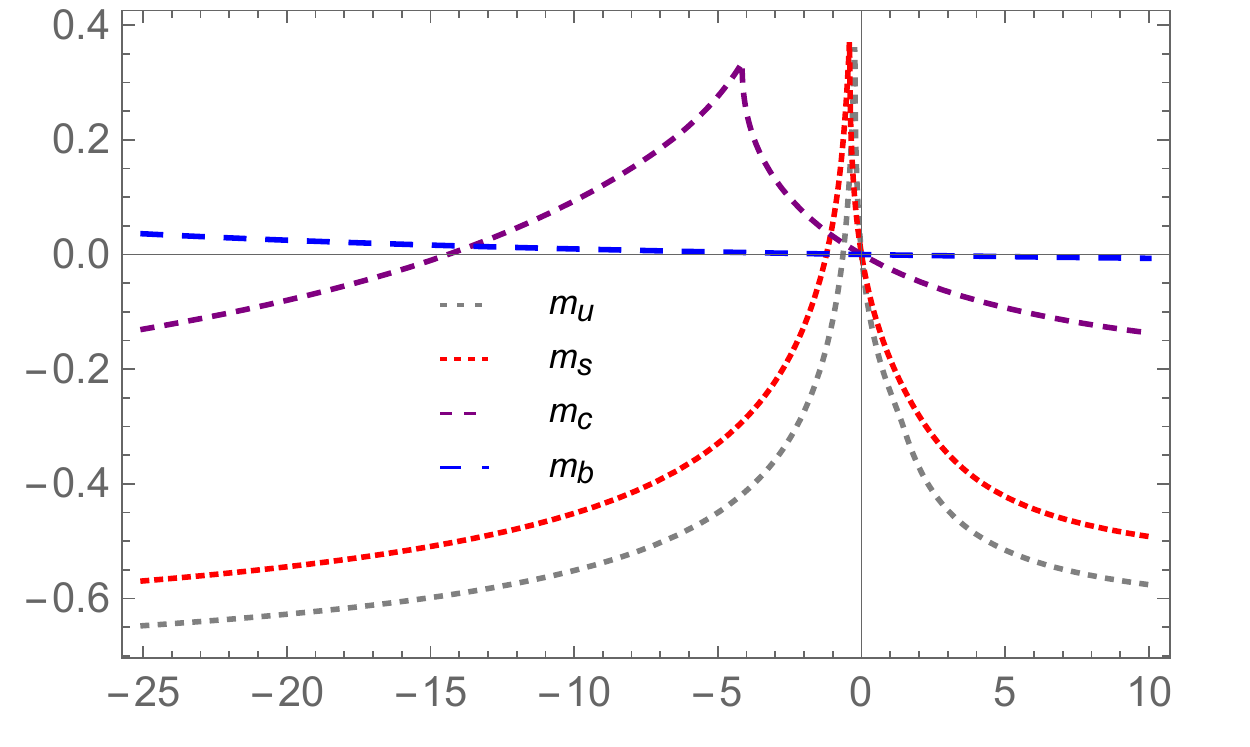}\vspace*{1ex}
  
  \includegraphics[width=\columnwidth]{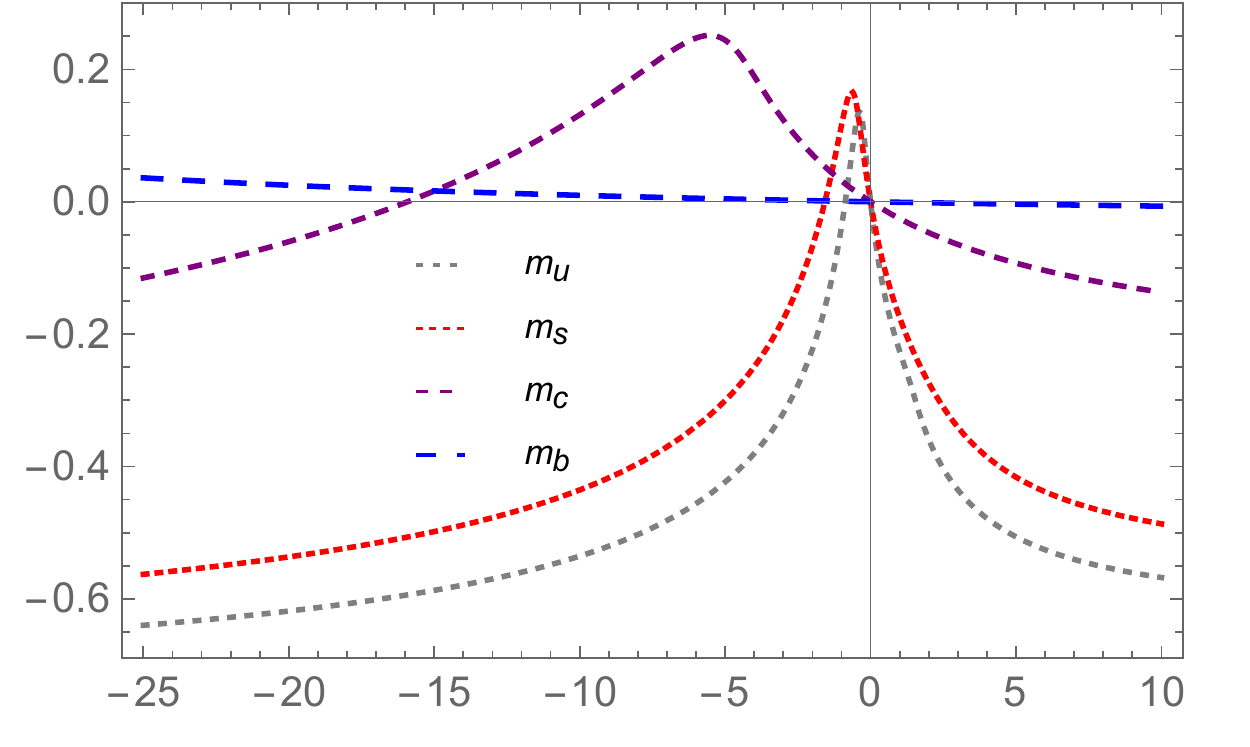}
  \includegraphics[width=\columnwidth]{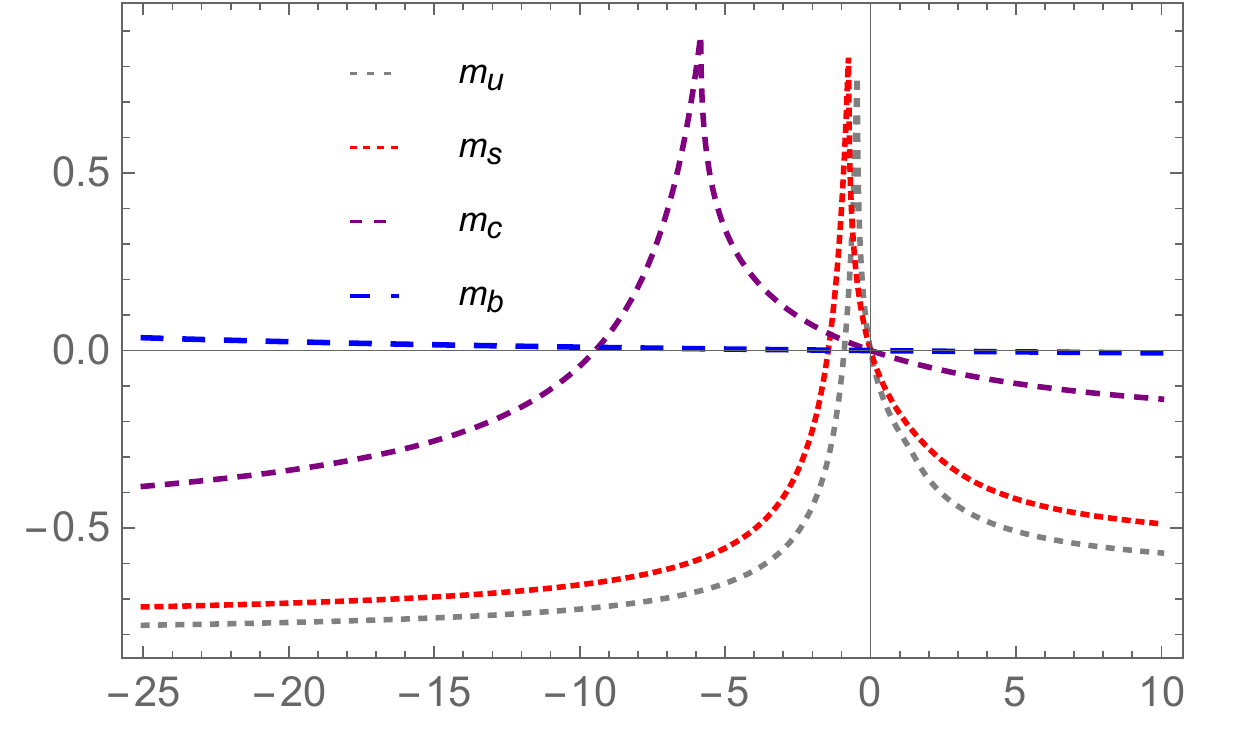}
\caption{\label{fig:arel}
Comparison of the relative difference of our calculated $A(s)$ to an \emph{Ansatz} $A(s)=1$ for the four quark masses, plotted on the same domain for each $n$. In particular, we plot $A(s)-1$. Left upper panel: $n=0$; right upper panel: $n=1$;
Left middle panel: $n=2$; right middle panel: $n=3$; Left lower panel: $n=4$; right lower panel: $n=\infty$.}
\end{figure*}

In this appendix we investigate and illustrate the properties of the quark-propagator
dressing functions $A$ and $M$ for heavy quarks. This allows us to provide support
for \emph{Ans\"atze} used for the quark propagator in the heavy-quark case, where
$A(s)$ is approximated by $1$ and $M(s)$ is approximated by a suitable value $M_q$,
both of which are assumed constant. 

To check this approximation we compare our calculated propagator dressing functions
to the assumed effective version with $A(s)=1$ and $M(s)=M(0)$. As a first illustration
we present $A(s)$ for all four quark flavors under consideration on the same domain,
namely from the relevant timelike point in bottomonium to the spacelike region. In Fig.~\ref{fig:adiff} we
plot the four curves corresponding to the four quark masses separately for each $n$
and find qualitatively interesting features: First of all, on the considered domain,
for all $n$, the behavior of $A(s)$ resembles that of a constant only for the $b$ quark;
the $c$-quark propagator is still considerably dressed. Secondly, while one has
to be aware of relevant domains for the lighter quarks regarding their respective bound-state-mass
defined scales, the first point remains valid also if one only looks at an
appropriately rescaled smaller portion of the timelike domain.

\begin{figure*}[t]
  \includegraphics[width=\columnwidth]{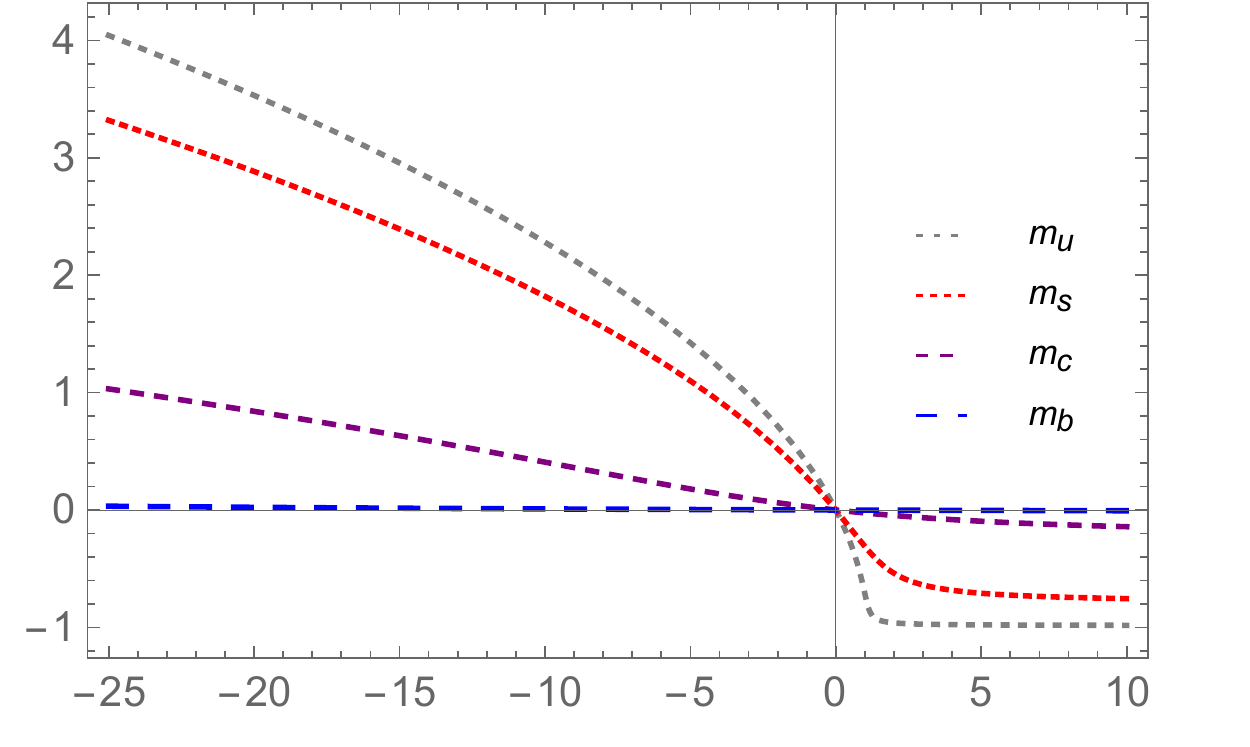}
  \includegraphics[width=\columnwidth]{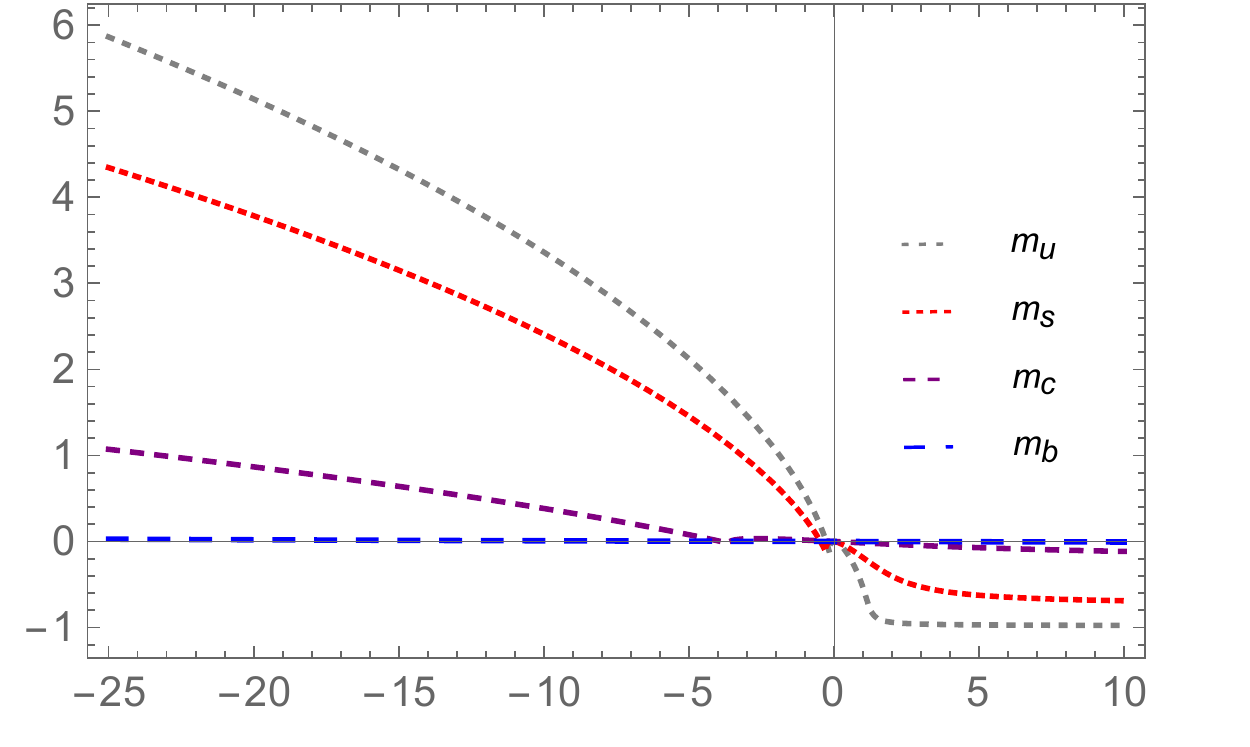}\vspace*{1ex}
  
  \includegraphics[width=\columnwidth]{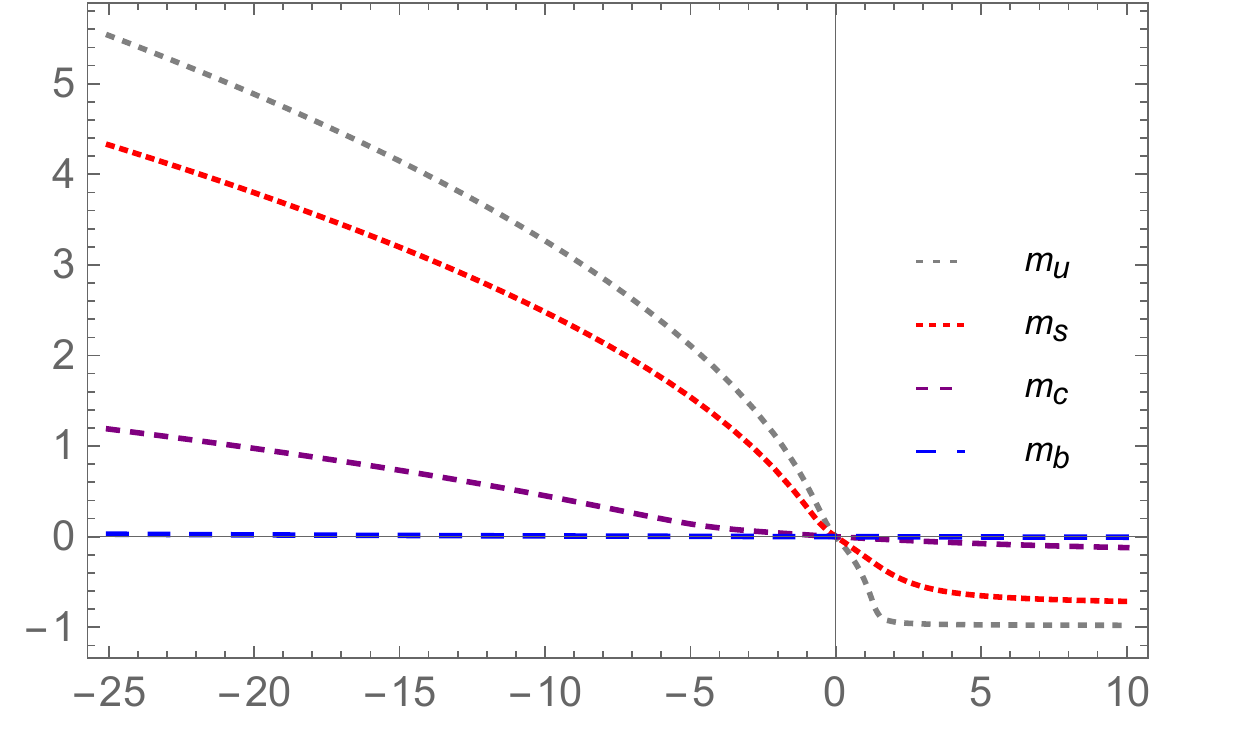}
  \includegraphics[width=\columnwidth]{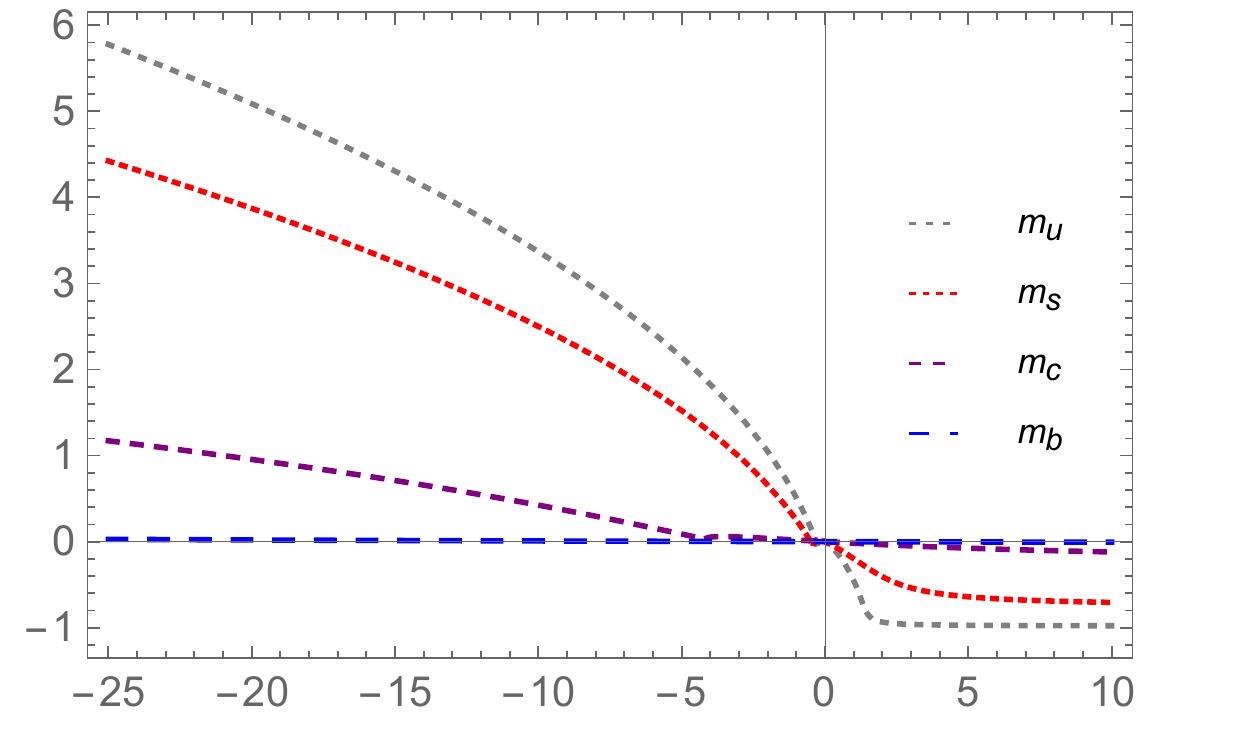}\vspace*{1ex}
  
  \includegraphics[width=\columnwidth]{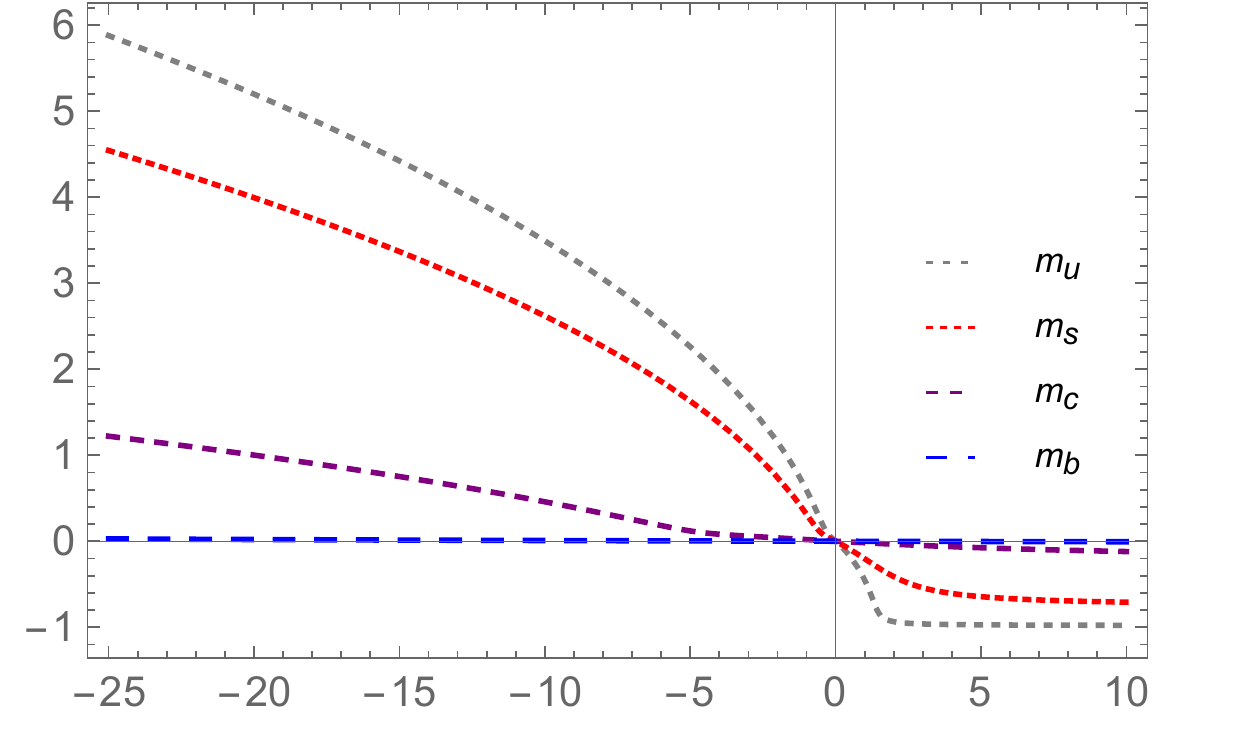}
  \includegraphics[width=\columnwidth]{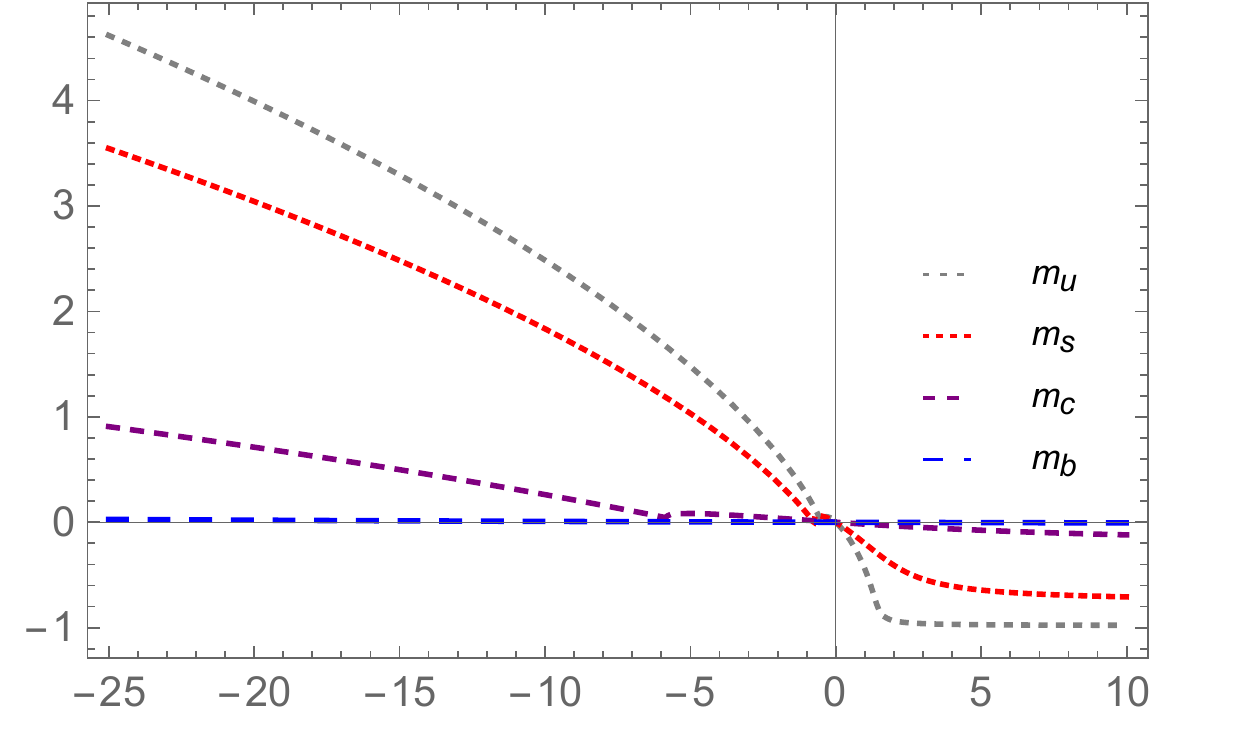}
\caption{\label{fig:mrel}
Comparison of the relative difference of our calculated $M(s)$ to an \emph{Ansatz} $M(s)=\mbox{const.}=M(s=0)$ for the four quark masses, plotted on the same domain for each $n$. In particular, we plot $[M(s)-M(0)]/M(0)$. Left upper panel: $n=0$; right upper panel: $n=1$;
Left middle panel: $n=2$; right middle panel: $n=3$; Left lower panel: $n=4$; right lower panel: $n=\infty$.}
\end{figure*}

As a next step, we compare both calculated dressing functions $A$ and $M$ to their effective constant counterparts
by plotting the relative difference between dressed and constant versions. The results for $A(s)$ are shown
in Fig.~\ref{fig:arel}, again each group of the four quark mass versions separately for the usual values of $n$.
The same is shown in Fig.~\ref{fig:mrel} for $M(s)$. For the constant effective value of $M_q$ we choose
$M(s=0)$, wich is one possible definition and for our purposes is as good as any other choice on that domain
of $s$ on which dynamical chiral symmetry breaking is clearly visible.

Apparently, our figures suggest that the transition from dressed to effective quark propagator happens
somewhere between the $c$ and $b$ quark masses, which is not entirely unexpected: charm quarks are in
general not regarded as heavy in the sense that effective theories work there without trouble; bottom
quarks, on the other hand, are usually regarded as well-approximated by an effective form of the quark
propagator, which we find as well. The key in Figs.~\ref{fig:arel} and \ref{fig:mrel} in order to decide
whether or not one is close to the form of a free propagator is the proximity
of the relative difference to zero and a constant type of behavior on the relevant domain. The only case
in our computed results where this is clearly visible, is the case of the $b$ quark.

\begin{figure*}[t]
  \includegraphics[width=0.9\columnwidth]{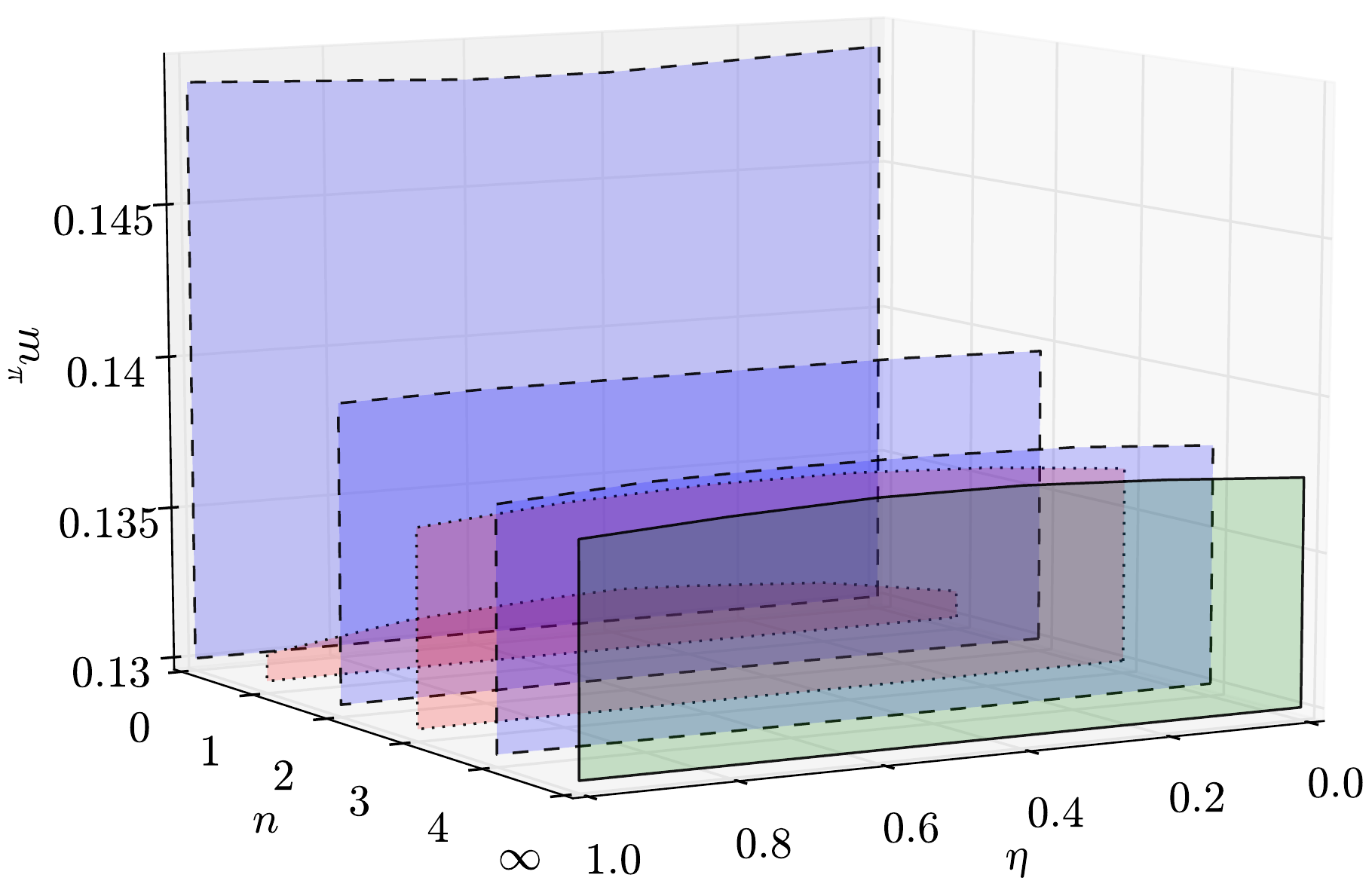}
  \includegraphics[width=0.9\columnwidth]{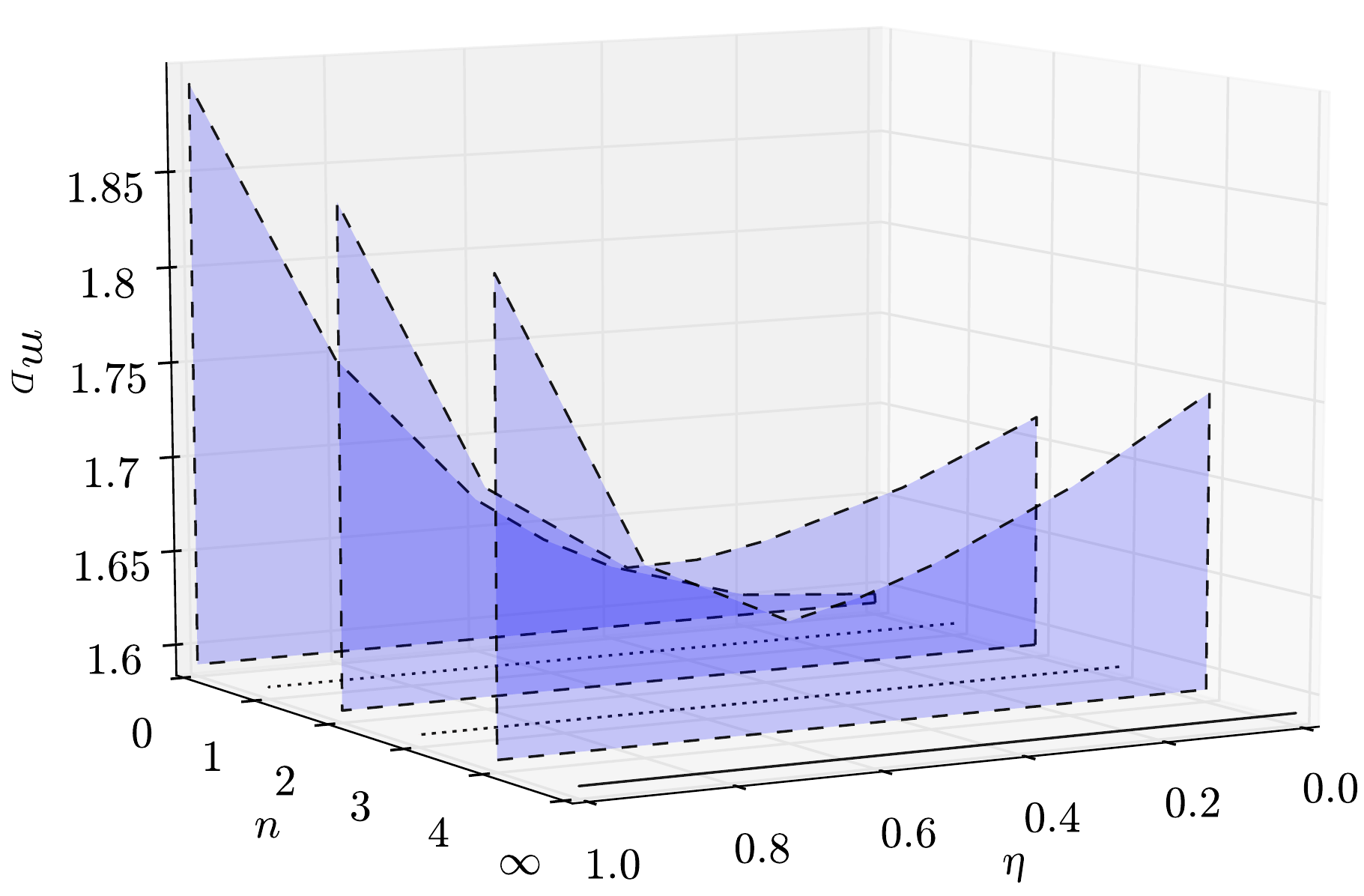}
  \includegraphics[width=0.9\columnwidth]{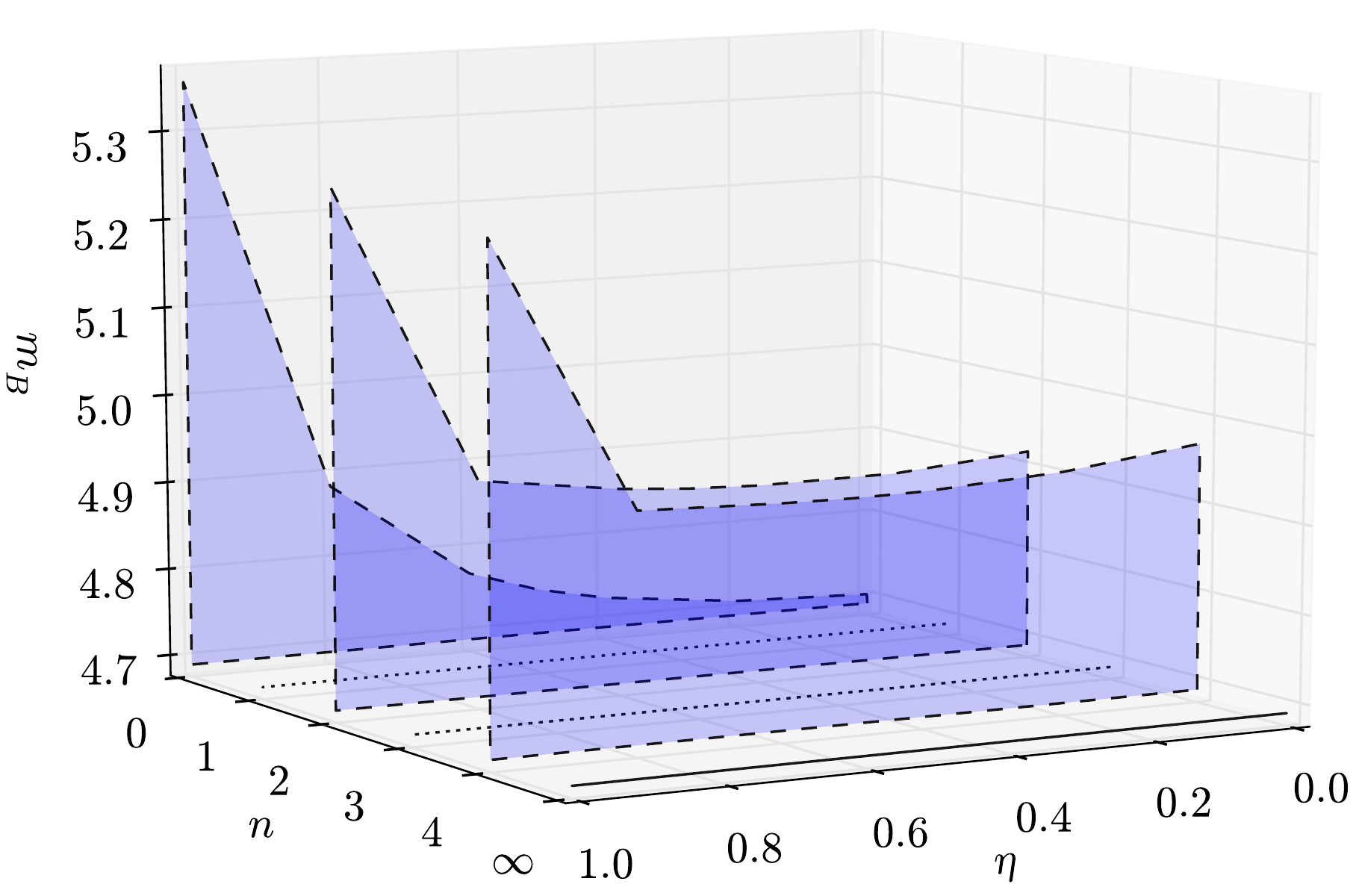}
  \includegraphics[width=0.9\columnwidth]{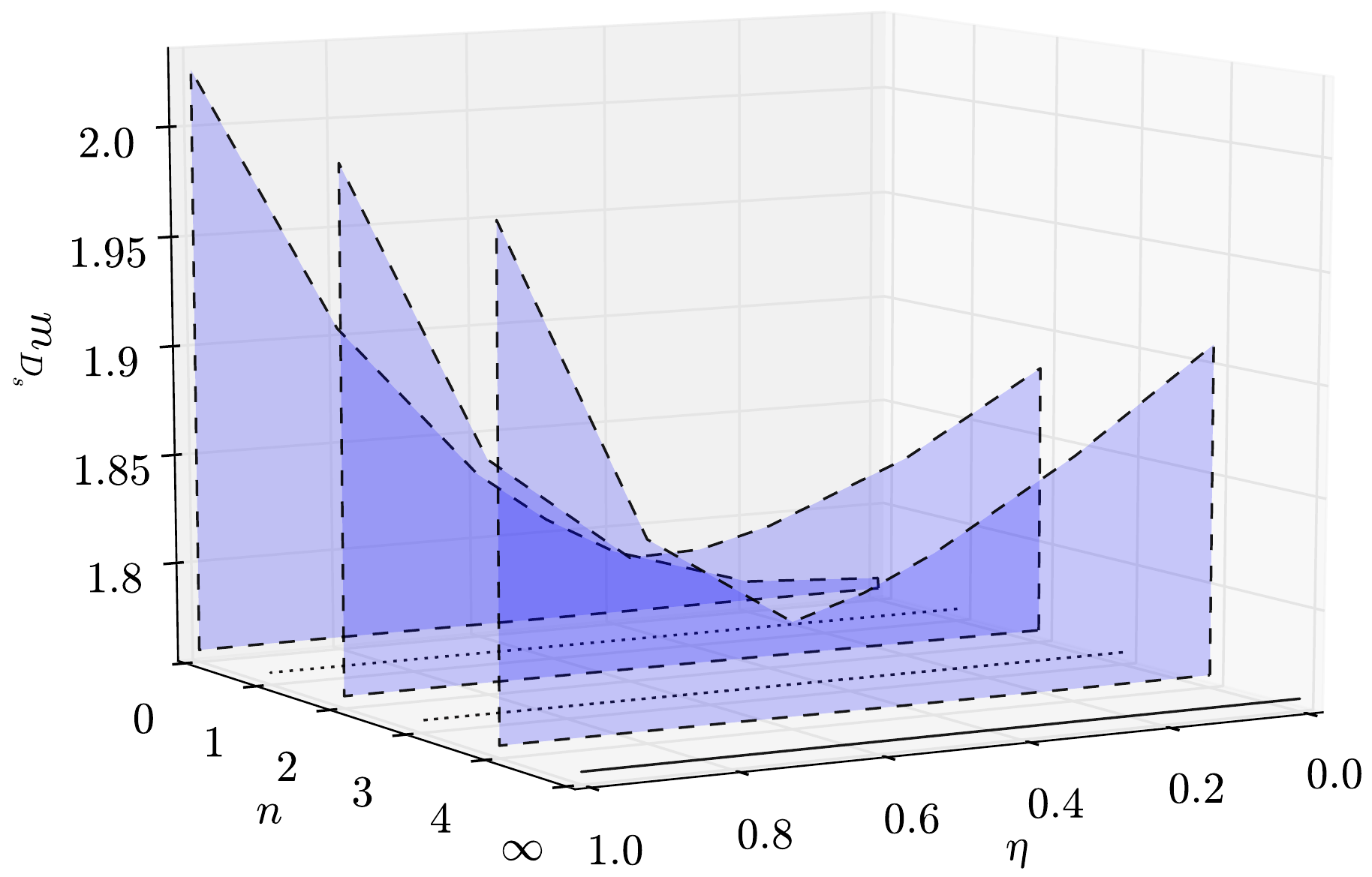}
  \includegraphics[width=0.9\columnwidth]{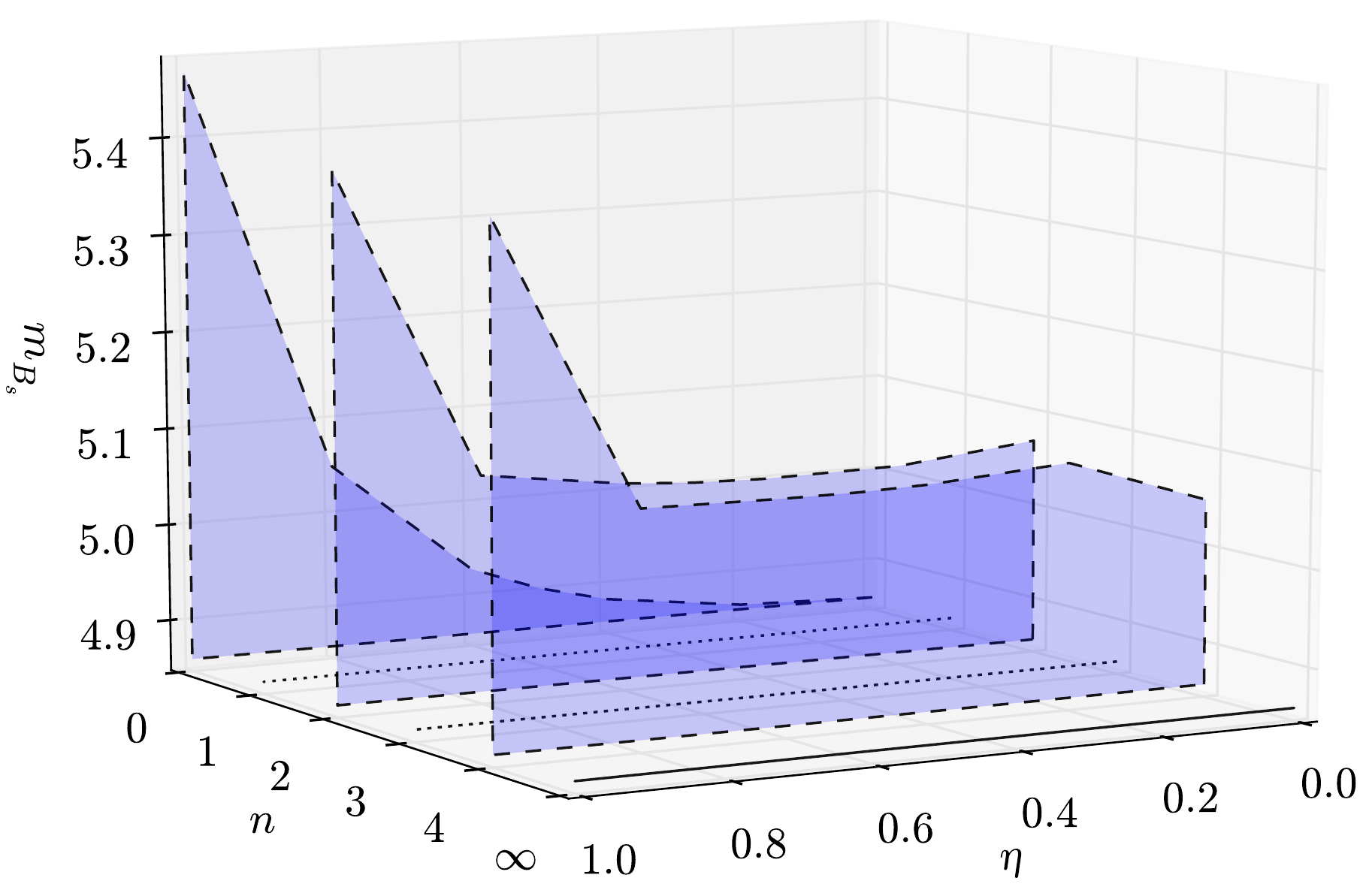}
  \includegraphics[width=0.9\columnwidth]{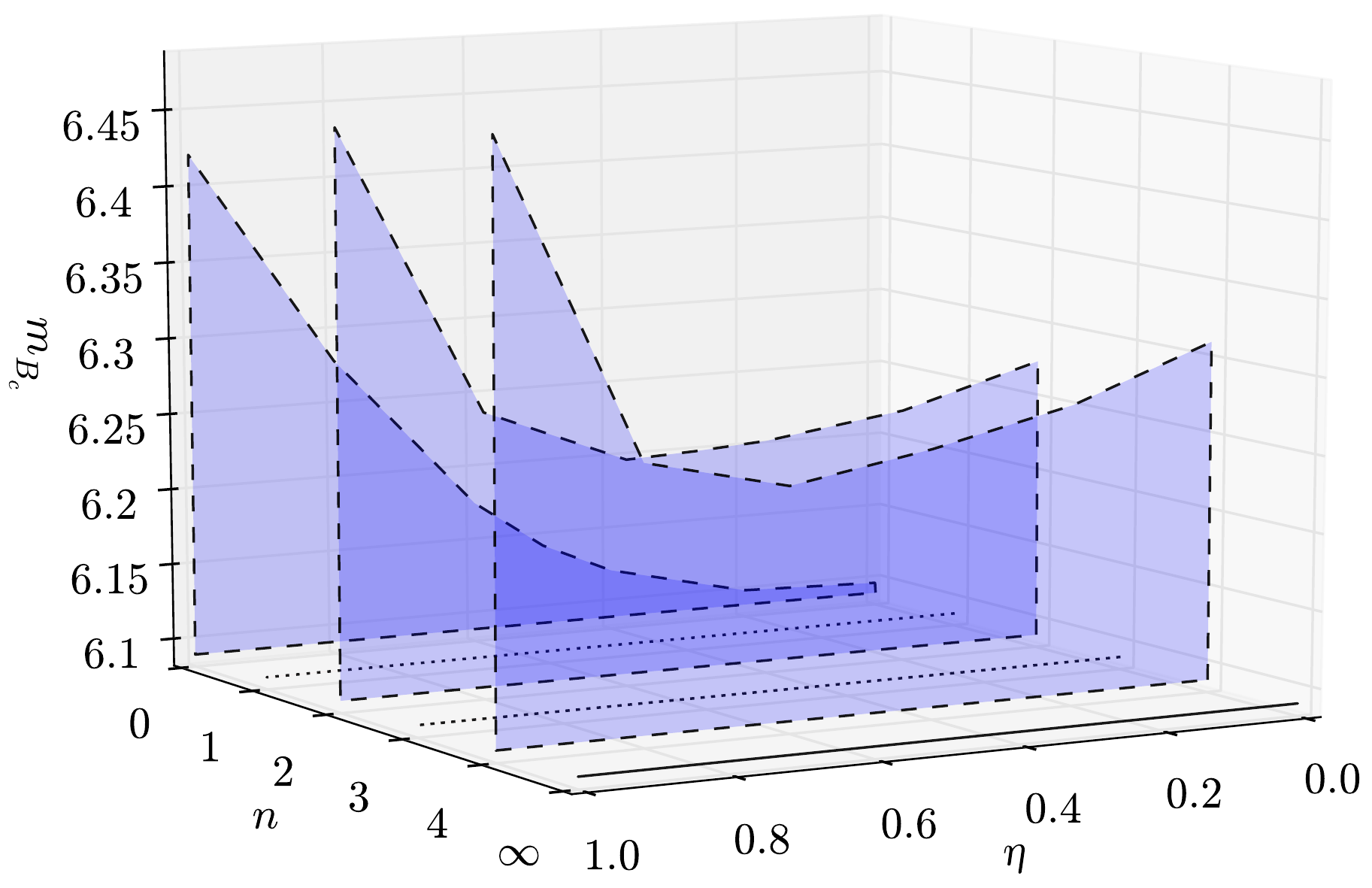}
  \includegraphics[width=0.9\columnwidth]{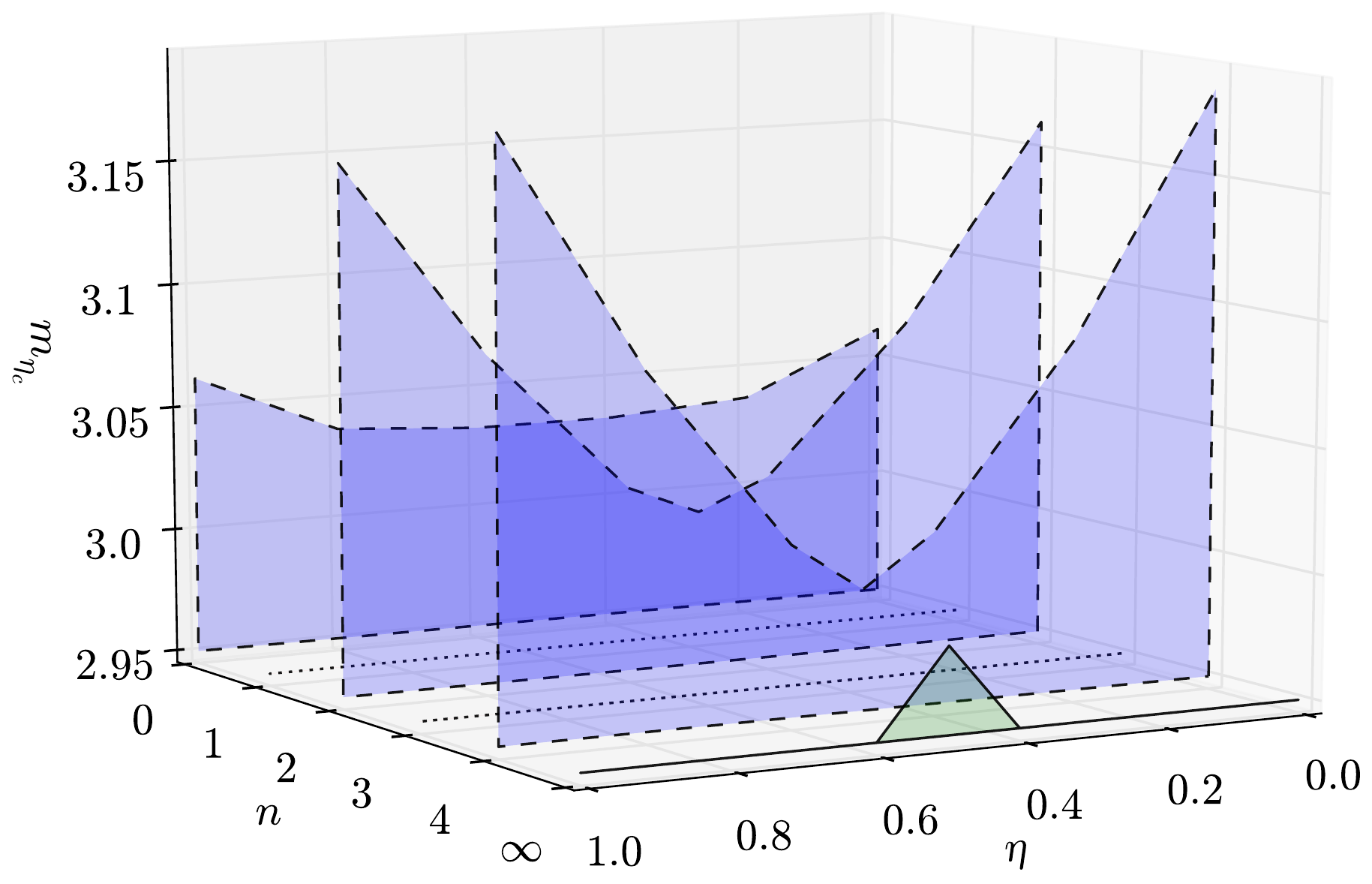}
  \includegraphics[width=0.9\columnwidth]{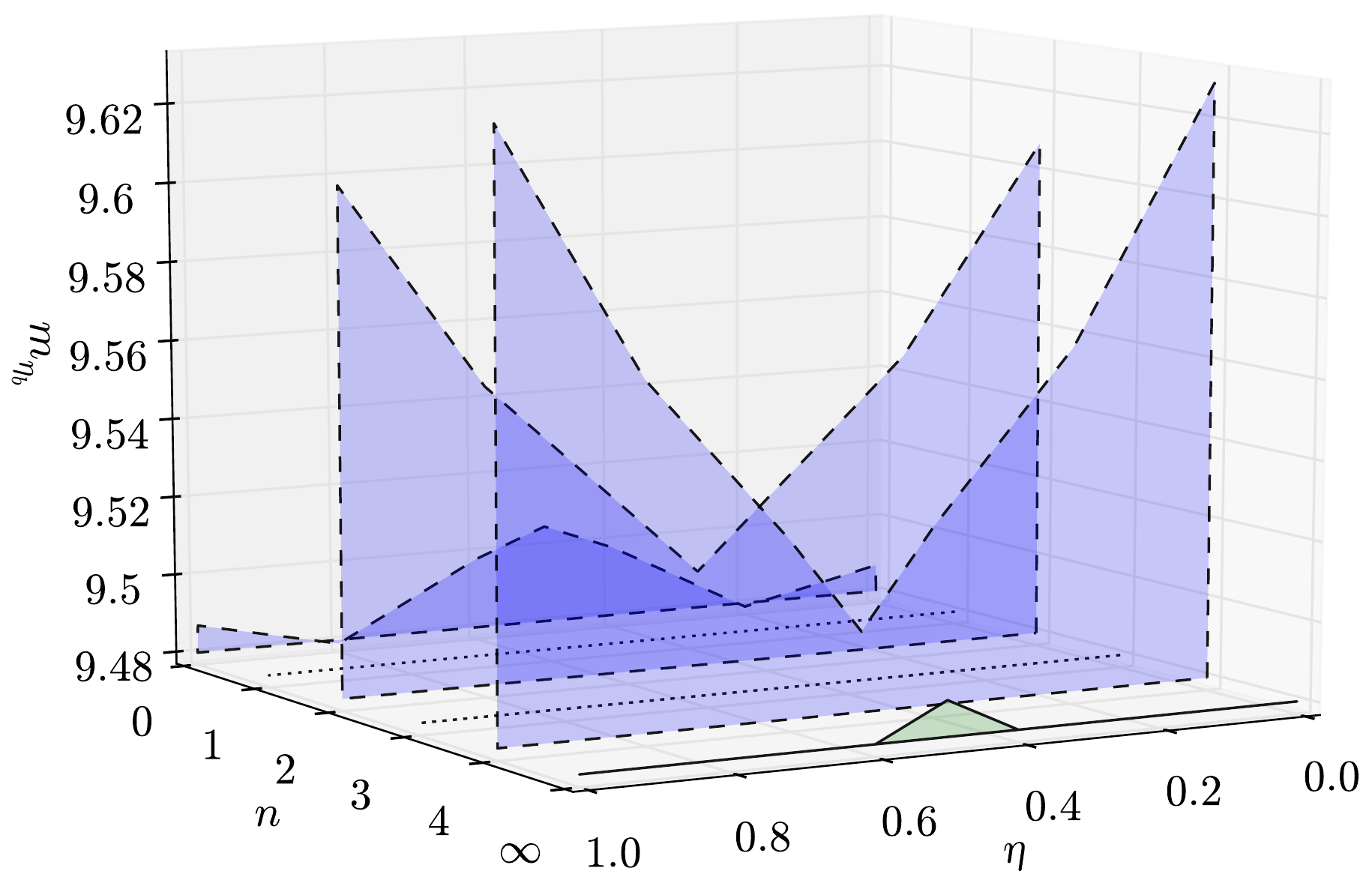}
\caption{\label{fig:mall}
Meson bound-state masses as functions of $n$ and $\eta$, given in GeV. Even $n$ are depicted by dashed
lines, odd ones by dotted lines, and the fully summed result by a solid line. If no solution is found,
no surface is plotted at the corresponding $n$. Left upper panel: pion; right upper panel: $D$;
Left upper center panel: $B$; right upper center panel: $D_s$; Left lower center panel: $B_s$; 
right lower center panel: $B_c$ Left lower panel: $\eta_c$; right lower panel: $\eta_b$.}
\end{figure*}

\section{$\eta$-dependence of meson masses}\label{sec:etadependence}
In this appendix we collect data and plots about the details of the
dependence of the meson masses on the momentum-partitioning parameter $\eta$
as a function of the order $n$ in our scheme in analogy to Tab.~\ref{tab:mkaon} and 
Fig.~\ref{fig:mkaon} in the main text of the paper. The data is collected in
Tab.~\ref{tab:mall}; the corresponding plots are shown
in the various panels of Fig.~\ref{fig:mall}. As a reference and excellent
example for the kind of behavior to expect, we present the pion, $\eta_c$, and $\eta_b$ cases
in addition to the unequal-mass collection of data and plots. For the pion
the alternating pattern of convergence of the odd and even $n$ numbers 
described above in connection with the kaon case is reobserved. For the other cases, 
the situation is somewhat obscured by the lack of solutions for odd $n$ on our
main $\eta$ grid points. Still, one can see convergence with $n$ as well as
a pronounced $\eta$ asymmetry for the heavy-light case, which is the source
of the large error bars plotted in Fig.~\ref{fig:expcomp}. Notwithstanding this,
the results clearly corroborate the systematic character of both the approach 
and the truncation scheme presented here as well as the validity of qualitative
as well as quantitative statements made above.

\begin{table*}[t]
\caption{Bound-state masses for various mesons the as functions of $n$ and $\eta$, given in GeV. \label{tab:mall}}
  \begin{tabular*}{0.98 \columnwidth}{@{\extracolsep{\fill} } c  c  c  c  c  c  c  c }\hline\hline
  						\multicolumn{8}{c}{$\pi$}\\
    			    $\eta$   &   0       &  $0.2$    &  $0.4$    &  $0.5$   &  $0.6$   &  $0.8$   &   1      \\ \hline
                            $n=0$    &  0.1490   &  0.1488   &  0.1486   &  0.1486  &  0.1486  &  0.1488  &  0.1490  \\
                            $n=1$    &  0.1309   &  0.1316   &  0.1319   &  0.1320  &  0.1319  &  0.1316  &  0.1309  \\
                            $n=2$    &  0.1397   &  0.1398   &  0.1398   &  0.1398  &  0.1398  &  0.1398  &  0.1397  \\
                            $n=3$    &  0.1364   &  0.1368   &  0.1370   &  0.1370  &  0.1370  &  0.1368  &  0.1364  \\  
                            $n=4$    &  0.1378   &  0.1381   &  0.1382   &  0.1382  &  0.1382  &  0.1381  &  0.1378  \\ \hline
  		        $n=\infty$   &  0.1374   &  0.1377   &  0.1379   &  0.1379  &  0.1379  &  0.1377  &  0.1374  \\\hline\hline
  \end{tabular*}\hfill
  \begin{tabular*}{0.98 \columnwidth}{@{\extracolsep{\fill} } c  c  c  c  c  c  c  c }\hline\hline
  						\multicolumn{8}{c}{$D$}\\
			    $\eta$   &  $0$ &  $0.2$ &  $0.4$  &  $0.5 $ &  $0.6$  & $0.8$  & 1  \\ \hline
                            $n=0$    &  1.645    &  1.651      &    1.673     &   1.691      &   1.716      &   1.797     & 1.946     \\
                            $n=1$    &  ... &   ...        &    ...       &   ...        &   ...        &     ...     & ...\\
                            $n=2$    &  1.763    &     1.731    &    1.709     &    1.702     &     1.701    &  1.749    & 1.900 \\
                            $n=3$    &  ... &     ...       &     ...      &  ...         &  ...         & ....            & ...  \\  
                            $n=4$    &  1.795    &    1.752     &    1.718     &    1.705     &    1.696     &    1.731       & 1.881  \\ \hline
  						    $n=\infty$   & ...  &  ...      &     ...      &   ...        &      ...     &  ...   & ... \\\hline\hline
  \end{tabular*}

\vspace*{3ex}

  \begin{tabular*}{0.98 \columnwidth}{@{\extracolsep{\fill} } c  c  c  c  c  c  c  c }\hline\hline
  						\multicolumn{8}{c}{$B$}\\
    			     $\eta$ &  $0$ &    $0.2$   &    $0.4$  &   $0.5 $   &  $0.6$   &  $0.8$ & 1    \\ \hline
                            $n=0$    &  4.801    &    4.807   &   4.825   &   4.842     &   4.868      &  4.983 &  5.456        \\
                            $n=1$    & ...  &     ...    &    ...    &   ...        &   ...        &     ... &  ... \\
                            $n=2$    &  5.017    &    5.004   &   5.003   &      5.006   &   5.012      &   5.034 &  5.370 \\
                            $n=3$    & ...  &    ...     &     ...    &  ...         &  ...         & ....       &   ...    \\  
                            $n=4$    & 5.069     &   5.050    &   5.041     &   5.040      &   5.041      &     5.046    &  5.350  \\ \hline
  						    $n=\infty$   &  ... &  ...      &     ...      &   ...        &      ...     &  ... & ... \\\hline\hline
  \end{tabular*}\hfill
  \begin{tabular*}{0.98 \columnwidth}{@{\extracolsep{\fill} } c  c  c  c  c  c  c  c }\hline\hline
  						\multicolumn{8}{c}{$D_s$}\\
    			     $\eta$  &   0 &    $0.2$ &  $0.4$      &   $0.5 $    &    $0.6$     &  $0.8$  & 1   \\ \hline
                            $n=0$    &  1.815    &    1.819 &   1.839     &    1.857    &    1.881     &   1.954     & 2.076     \\
                            $n=1$    &  ... &   ...    &    ...      &   ...       &   ...        &     ...    & ... \\
                            $n=2$    &  1.933     &  1.896  &   1.870     &    1.862    &   1.861      &   1.911    & 2.047 \\
                            $n=3$    &  ...  &   ...   &     ...     &  ...        &  ...         &    ....      &   ...      \\  
                            $n=4$    &  1.960     &  1.915  &   1.877     &  1.862      &   1.852      &   1.894       &  2.036  \\ \hline
  						    $n=\infty$   & ...     &  ...      &     ...      &   ...        &      ...     &  1.880  & ...\\\hline\hline
  \end{tabular*}

\vspace*{3ex}

  \begin{tabular*}{0.98 \columnwidth}{@{\extracolsep{\fill} } c  c  c  c  c  c  c  c }\hline\hline
  						\multicolumn{8}{c}{$B_s$}\\
    			     $\eta$  &  0      &  $0.2$   &  $0.4$      &   $0.5 $    &    $0.6$     &  $0.8$    & 1      \\ \hline
                            $n=0$    &  4.960  &   4.965  &   4.984     &   5.002     &   5.029      &  5.150    &  5.567 \\
                            $n=1$    &  ...    &   ...    &    ...      &   ...       &   ...        &   ...     &  ...   \\
                            $n=2$    &  5.171  &   5.156  &   5.154     &   5.156     &   5.161      &  5.181    &  5.497 \\
                            $n=3$    &  ...    &    ...   &    ...      &   ...       &   ...        &   ....    &  ...   \\  
                            $n=4$    &  5.150  &   5.199  &   5.189     &   5.187     &   5.187      &  5.189    &  5.483 \\ \hline
  						    $n=\infty$   &  ...  &  ...       &     ...      &   ...     &      ...     &  ...  &  ... \\\hline\hline
  \end{tabular*}\hfill
  \begin{tabular*}{0.98 \columnwidth}{@{\extracolsep{\fill} } c  c  c  c  c  c  c  c }\hline\hline
  						\multicolumn{8}{c}{$B_c$}\\
    			    $\eta$   &   0      &  $0.2$   &  $0.4$    &   $0.5 $    &   $0.6$    &  $0.8$      &  1      \\ \hline
                            $n=0$    &  6.147   &  6.150   &  6.172    &   6.193     &  6.226     &  6.329      &  6.471  \\
                            $n=1$    &  ...     &   ...    &    ...    &   ...       &   ...      &     ...     &  ...    \\
                            $n=2$    &  6.325   &  6.299   &  6.286    &   6.283     &  6.281     &  6.319      &  6.507  \\
                            $n=3$    &  ...     &   ...    &    ...    &   ...       &   ...      &    ....     &  ...    \\  
                            $n=4$    &  6.361   &  6.327   &  6.306    &   6.298     &  6.290     &  6.312      &  6.521  \\ \hline
  						    $n=\infty$   &  ...  &  ...      &   ...      &   ...       &   ...     &  ...  &  ... \\ \hline\hline
  \end{tabular*}

\vspace*{3ex}

  \begin{tabular*}{0.98 \columnwidth}{@{\extracolsep{\fill} } c  c  c  c  c  c  c  c }\hline\hline
  						\multicolumn{8}{c}{$\eta_c$}\\
    			     $\eta$  &  0      &  $0.2$   &  $0.4$      &   $0.5 $    &    $0.6$     &  $0.8$    & 1      \\ \hline
                            $n=0$    &  3.062    &  3.037   &   3.033  &   3.033  &  3.033  &  3.037   &  3.062  \\
                            $n=1$    &    ...    &   ...    &   ...    &   ...    &   ...   &   ...    &   ...   \\
                            $n=2$    &  3.161    &  3.082   &   3.024  &   3.012  &  3.024  &  3.082   &  3.161  \\
                            $n=3$    &    ...    &   ...    &   ...    &   ...    &   ...   &   ...    &   ...   \\  
                            $n=4$    &  3.185    &  3.090   &   3.018  &   2.998  &  3.018  &  3.090   &  3.185  \\ \hline
  						    $n=\infty$   &  ...  &  ...       &     ...      &   2.985     &      ...     &  ...  &  ... \\\hline\hline
  \end{tabular*}\hfill
  \begin{tabular*}{0.98 \columnwidth}{@{\extracolsep{\fill} } c  c  c  c  c  c  c  c }\hline\hline
  						\multicolumn{8}{c}{$\eta_b$}\\
    			    $\eta$   &   0      &  $0.2$   &  $0.4$    &   $0.5 $    &   $0.6$    &  $0.8$      &  1      \\ \hline
                            $n=0$    &   9.487   &  9.479   &  9.498   &  9.505   &  9.498  &  9.479   &   9.487  \\
                            $n=1$    &    ...    &   ...    &   ...    &   ...    &   ...   &   ...    &    ...   \\
                            $n=2$    &   9.607   &  9.555   &  9.521   &  9.504   &  9.521  &  9.555   &   9.607  \\
                            $n=3$    &    ...    &   ...    &   ...    &   ...    &   ...   &   ...    &    ...   \\  
                            $n=4$    &   9.629   &  9.566   &  9.524   &  9.500   &  9.524  &  9.566   &  9.629   \\ \hline
  						    $n=\infty$   &  ...  &  ...      &   ...      &   9.489       &   ...     &  ...  &  ... \\ \hline\hline
  \end{tabular*}
\end{table*}

\section{pseudoscalar kernel details}\label{sec:pskerneldetails}

Following Refs.~\cite{Bender:2002as,Bhagwat:2004hn} we start from the recursion relations 
for the QGV $\Gamma_\mu$, Eq.~(\ref{eq:qgvrecursion}) and the
BSE correction term $\Lambda_{M\mu}$, Eq.~(\ref{eq:recbiglambda}). The
first step in our construction is the decomposition of each of these
quantities in terms of Dirac covariants. 

The full QGV has 12 covariant structures built from $\gamma_\mu$ and the two
independent four-vectors of the quark and antiquark lines. After
the effective interaction, Eq.~(\ref{eq:mnmodel}) has been employed, only
three of those are nonzero and one obtains
\begin{equation}\label{eq:qgvstructure}
\Gamma_\mu(p)=\alpha_1(p^2)\gamma_\mu + \alpha_2(p^2)\gamma\cdot p\; p_\mu - i \alpha_3(p^2) p_\mu\;.
\end{equation}
With this structure and the initial condition that the QGV be bare
\begin{equation}\label{eq:qgvinitial}
\Gamma_\mu(p)^0=\gamma_\mu 
\end{equation}
one can express the QGV via its recursion relation to any desired order 
in terms of the covariants given in Eq.~(\ref{eq:qgvstructure}) and the quark propagator $S(p)$. The result, in turn, can be inserted
in the quark DSE and yields algebraic equations for $A(s)$ and $M(s)$ via Dirac-trace projections
onto the two covariant quark propagator structures.

A similar strategy is used to compute $\Lambda_{M\mu}(P)$. One starts off with finding a suitable
decomposition in terms of Dirac covariants for the quantum numbers appropriate for the meson $M$ under
consideration, in our case pseudoscalar. In our setup the pseudoscalar BSA has 2 nonzero components
from the general four:
\begin{equation}\label{eq:psbsa}
\Gamma_{PS}(P)=f_1(P^2)\,i\;\gamma_5 +  f_2(P^2)\,\gamma_5\; \gamma\cdot \hat{P}
\end{equation}
with the unit vector $\hat{P}:=P/\sqrt{P^2}$.

The corresponding $\Lambda_{0^{-}\nu}$ has in general 12 covariant
structures, four of which are nonzero in our particular setup. We have, omitting the subscript 
denoting the pseudoscalar case
\begin{eqnarray}\nonumber
\Lambda_\mu(P)&=&\beta_1(P^2)\,\gamma_5 \gamma_\mu + \beta_2(P^2)\,\gamma_5\;\gamma\cdot \hat{P}\; \hat{P}_\mu  
	\\ \label{eq:lambdastructure}
	&+&  \beta_3(P^2) \,\gamma_5 \hat{P}_\mu +  \beta_4(P^2)\, \gamma_5\;\gamma_\mu\; \gamma\cdot \hat{P}\; .
\end{eqnarray}

One can obtain the scalar functions $\vec{\beta}:=\{\beta_j\}$, $j=1,2,3,4$, from a recursion relation extracted
from the recursion relations for the QGV and $\Lambda$ via defining the projection operators
\begin{eqnarray}
\mathcal{P}_{\mu,1}&:=&\frac{1}{16}\gamma_\mu\;\gamma_5  \\
\mathcal{P}_{\mu,2}&:=&\frac{1}{4}\hat{P}_\mu\;\gamma\cdot \hat{P}\;\gamma_5  \\
\mathcal{P}_{\mu,3}&:=&\frac{1}{4}\hat{P}_\mu\;\gamma_5  \\
\mathcal{P}_{\mu,4}&:=&\frac{1}{16}\cdot \hat{P}\;\gamma_\mu\;\gamma_5  \;.
\end{eqnarray}
Then at every $i$ one has
\begin{equation} \label{eq:betamatrix} 
\beta_j^i = (\mathcal{M}_{\text{ps}})_{jk} \mathrm{Tr}\left[\mathcal{P}_{\mu,k} \Lambda^i_{\mu,\text{ps}} \right] \;,
\end{equation} 
where the matrix $\mathcal{M}_{\text{ps}}$ is given by
\begin{equation}
\mathcal{M}_{\text{ps}}=
\left(
\begin{array}{cccc}
 \frac{4}{3} & -\frac{1}{3} & 0 & 0 \\
 -\frac{4}{3} & \frac{4}{3} & 0 & 0 \\
 0 & 0 & \frac{4}{3} & -\frac{4}{3} \\
 0 & 0 & -\frac{1}{3} & \frac{4}{3} \\
\end{array}
\right)
\end{equation}
and one can write (see also Ref.~\cite{Bender:2002as} for more details)
\begin{equation} 
\vec{\beta}^i = \mathcal{M}_{\text{ps}} \vec{\mathcal{T}}^i\;.
\end{equation} 
The vector $\vec{\mathcal{T}}^i$, in turn is obtained, writing $\vec{\alpha}:=\{\alpha_j\}$, $j=1,2,3$, from 
\begin{equation} 
\vec{\mathcal{T}}^i  = \mathbf{G}_-\,\vec{\alpha}_-^{\,i-1}  +\mathbf{G}_+\,\vec{\alpha}_+^{\,i-1} + 
\mathbf{L}\, \vec{\beta}^{\,i-1}, 
\label{distributive} 
\end{equation} 
which can be calculated when the matrices $\mathbf{G}$ and $\mathbf{L}$ are known.
$\vec{\alpha}_+$ and $\vec{\alpha}_-$ denote the coefficients of the QGV decomposition
corresponding to the ${}_+$ and  ${}_-$ arguments appearing in their defining quark 
propagators as given above.
Similarly, we define
\begin{eqnarray}
 B_- &:=&  M(p_-^2) A(p_-^2) \\
 B_+ &:=&  M(p_+^2) A(p_+^2) \\
 A_- &:=&  (\eta -1)\,\sqrt{P^2} A(p_-^2) \\
 A_+ &:=&  \eta \, \sqrt{P^2}   A(p_+^2) 
\end{eqnarray}
\noindent and
\begin{eqnarray}
\Delta_- &:=& B^2 (p_-^2) + P^2 A^2(p_-^2)   \\
\Delta_+ &:=&  B^2 (p_+^2) + P^2 A^2(p_+^2)
\end{eqnarray}
and obtain
\begin{widetext}
\begin{equation}
\mathbf{L}=
\frac{-\mathcal{C}}{\Delta_- \, \Delta_+ }
 \left(\begin{array}{cccc}
2 B_- B_+ - A_- A_+                               &  \frac{1}{2}\left( B_- B_+ + A_- A_+ \right)   &  -\frac{1}{2} i\left( A_- B_+  -A_+ B_- \right) & -i\left( 2A_- B_+ +A_+ B_- \right)    \\2\left( B_- B_+ + A_- A_+  \right)                &  2 \left( B_- B_+ + A_- A_+ \right)             &  -2 i  \left( A_- B_+-A_+ B_- \right)       &  -2 i  \left( A_- B_+  -A_+B_- \right)  \\
4 i   \left( A_- B_+  -A_+ B_- \right)  &  4 i \left( A_- B_+ - A_+B_-\right)    &  -4 \left( B_- B_+ + A_- A_+ \right)               &  -4 \left( B_- B_+ + A_- A_+ \right) \\
      i \left( A_- B_+  -A_+ B_-\right) &  i \left( A_- B_+-A_+B_- \right)         &  -(B_- B_+  +A_- A_+ )                                 &  -(B_- B_+ + A_- A_+)  
\end{array}
\right)\;.
\end{equation}
The two matrices $\mathbf{G}_-$ and $\mathbf{G}_+$ are associated with the corresponding quark propagators with the 
${}_+$ and  ${}_-$ arguments as defined above:


\begin{eqnarray}\nonumber
\mathbf{G}_-&=&\frac{-\mathcal{C}}{\Delta_-^2 \Delta_+}
\left(
\begin{array}{c}
 i \left(B_+ A_-^2+B_- A_+ A_-+2 B_-^2 B_+\right) f_1+i \left(A_+ A_-^2-B_- B_+ A_-+2 B_-^2 A_+\right) f_2 \\
 2 i \left(-B_+ A_-^2+2 B_- A_+ A_-+B_-^2 B_+\right) f_1+2 i \left(-A_+ A_-^2-2 B_- B_+ A_-+B_-^2 A_+\right) f_2 \\ 
 -4 \left(A_+ A_-^2+2 B_- B_+ A_--B_-^2 A_+\right) f_1-4 \left(-B_+ A_-^2+2 B_- A_+ A_-+B_-^2 B_+\right) f_2 \\ 
 \left(-A_+ A_-^2-2 B_- B_+ A_-+B_-^2 A_+\right) f_1+\left(B_+ A_-^2-2 B_- A_+ A_--B_-^2 B_+\right) f_2  
\end{array}\right. \\[2ex] \nonumber 
&&\begin{array}{c} 
 -\frac{1}{2} i
   P^2 \left(B_+ A_-^2-2 B_- A_+ A_--B_-^2 B_+\right) f_1 (\eta -1)^2-\frac{1}{2} i P^2 \left(A_+ A_-^2+2 B_- B_+ A_--B_-^2
   A_+\right) f_2 (\eta -1)^2 \\
 -2 i P^2
   \left(B_+ A_-^2-2 B_- A_+ A_--B_-^2 B_+\right) f_1 (\eta -1)^2-2 i P^2 \left(A_+ A_-^2+2 B_- B_+ A_--B_-^2 A_+\right)
   f_2 (\eta -1)^2 \\
 -4 P^2
   \left(A_+ A_-^2+2 B_- B_+ A_--B_-^2 A_+\right) f_1 (\eta -1)^2-4 P^2 \left(-B_+ A_-^2+2 B_- A_+ A_-+B_-^2 B_+\right) f_2
   (\eta -1)^2 \\ 
 -P^2 \left(A_+
   A_-^2+2 B_- B_+ A_--B_-^2 A_+\right) f_1 (\eta -1)^2-P^2 \left(-B_+ A_-^2+2 B_- A_+ A_-+B_-^2 B_+\right) f_2 (\eta -1)^2   
\end{array} \\[2ex] \nonumber 
&&\left.\begin{array}{c} 
   -\frac{1}{2} i \sqrt{P^2} (\eta -1) \left(A_+ A_-^2+2 B_- B_+ A_--B_-^2 A_+\right)
   f_1-\frac{1}{2} i \sqrt{P^2} (\eta -1) \left(-B_+ A_-^2+2 B_- A_+ A_-+B_-^2 B_+\right) f_2 \\
   -2 i \sqrt{P^2} (\eta -1) \left(A_+ A_-^2+2 B_- B_+ A_--B_-^2 A_+\right) f_1-2 i \sqrt{P^2} (\eta -1)
   \left(-B_+ A_-^2+2 B_- A_+ A_-+B_-^2 B_+\right) f_2 \\
   4 \sqrt{P^2} (\eta -1) \left(B_+ A_-^2-2 B_- A_+ A_--B_-^2 B_+\right) f_1+4 \sqrt{P^2} (\eta -1) \left(A_+
   A_-^2+2 B_- B_+ A_--B_-^2 A_+\right) f_2 \\
   \sqrt{P^2} (\eta -1) \left(B_+ A_-^2-2 B_- A_+ A_--B_-^2 B_+\right) f_1+\sqrt{P^2} (\eta -1) \left(A_+ A_-^2+2 B_- B_+
   A_--B_-^2 A_+\right) f_2    
\end{array}
\right)
\end{eqnarray}
which is to be understood as a $4\times 3$ matrix,
and its corresponding analog.
\end{widetext}

%
%
%


%

\end{document}